\documentclass[sigconf,nonacm]{acmart} 

\AtBeginDocument{}

\setcopyright{none} \copyrightyear{2018} \acmYear{2018} \acmDOI{XXXXXXX.XXXXXXX} \acmConference[Conference acronym 'XX]{Make sure to enter the correct
conference title from your rights confirmation email}{June 03--05,
2018}{Woodstock, NY} \acmISBN{978-1-4503-XXXX-X/2018/06} 

\makeatletter
\def\@ACM@checkaffil{}
\makeatother

\usepackage{tikz}
\usepackage{tabularx}
\usepackage{makecell}
\usepackage{listings}
\usepackage[caption=false,font=footnotesize]{subfig}
\usepackage[normalem]{ulem}
\usepackage{multirow}
\usepackage{algorithm}
\usepackage{algpseudocode}
\usepackage{enumitem}
\usepackage{xspace}
\usepackage{mathtools}
\usepackage{wasysym}
\usepackage[most]{tcolorbox}

\def\UrlBreaks{\do\/\do-}

\usepackage[textwidth=0.65in,textsize=scriptsize,colorinlistoftodos]{todonotes}
\let\pkgtodo\todo
\let\todo\relax

\definecolor{mGreen}{rgb}{0,0.6,0}
\definecolor{mGray}{rgb}{0.5,0.5,0.5}
\definecolor{mPurple}{rgb}{0.58,0,0.82}
\definecolor{backgroundColour}{rgb}{1,1,1}
\definecolor{lightgreen}{rgb}{0.8,1.0,0.8}
\definecolor{darkgreen}{rgb}{0.0, 0.5, 0.0}
\definecolor{lightblue}{rgb}{0.68, 0.85, 0.9}
\definecolor{lightpurple}{rgb}{0.7, 0.55, 0.88}

\definecolor{diffstart}{named}{blue}
\definecolor{diffrem}{named}{red}
\definecolor{black}{named}{black}

\lstdefinestyle{CStyleSmall}{
  backgroundcolor=\color{backgroundColour},
  commentstyle=\color{mGreen},
  keywordstyle=\color{mGray},
  numberstyle=\tiny\color{mGray},
  stringstyle=\color{black},
  basicstyle=\scriptsize\sffamily,
  breakatwhitespace=false,
  breaklines=true,
  captionpos=b,
  keepspaces=true,
  numbers=left,
  numbersep=5pt,
  showspaces=false,
  showstringspaces=false,
  showtabs=false,
  tabsize=2,
  breakindent=0pt,
moredelim=**[is][\color{diffrem}]{\\-}{\\-},
  moredelim=**[is][\color{darkgreen}]{\\+}{\\+},
  moredelim=**[is][\color{mGray}]{\\gray}{\\gray},
  moredelim=**[is][\color{mPurple}]{\\purple}{\\purple},
  moredelim=**[is][\bfseries]{\\b}{\\b},
  moredelim=**[is][\mdseries]{\\ub}{\\ub},
  moredelim=**[is][\color{black}]{\\uc}{\\uc},
}

\lstdefinestyle{diff}{
  basicstyle=\footnotesize\ttfamily,
  backgroundcolor=\color{backgroundColour},
  breakatwhitespace=false,
  breaklines=true,
  captionpos=b,
  keepspaces=true,
  numbers=left,
  numbersep=5pt,
  showspaces=false,
  showstringspaces=false,
  showtabs=false,
  tabsize=2,
  morecomment=[f][\color{diffstart}]{@@},
  morecomment=[f][\color{darkgreen}]{+},
  morecomment=[f][\color{diffrem}]{-},
}

\newlist{compactenum}{enumerate}{1}
\setlist[compactenum,1]{label=\arabic*., nosep}

\newlist{compactlist}{itemize}{3}
\setlist[compactlist]{label=\textbullet, nosep}

\newif\ifpresentMode
\presentModefalse 

\ifpresentMode
\newcommand{\todo}[1]{}
\else
\newcommand{\todo}[1]{\noindent\textcolor{red}{TODO: #1} \\}
\fi
\ifpresentMode
\newcommand{\mnote}[2][]{}
\newcommand{\mnoteyellow}[1]{}
\newcommand{\mnotered}[1]{}
\else
\newcommand{\mnote}[2][yellow]{\pkgtodo[color=#1!40,size=\scriptsize,linecolor=#1!60]{#2}}
\newcommand{\mnoteyellow}[1]{\mnote[yellow]{#1}}
\newcommand{\mnotered}[1]{\mnote[red]{#1}}
\fi
\ifpresentMode
\newcommand{\cv}[1]{}
\newcommand{\alp}[1]{}
\newcommand{\raluca}[1]{}

\else
\newcommand{\cv}[1]{\noindent\textcolor{blue}{Corban: #1}}
\newcommand{\alp}[1]{\noindent\textcolor{mPurple}{Alp: #1}}
\newcommand{\raluca}[1]{\noindent\textcolor{lightpurple}{Raluca: #1}}

\fi

\ifpresentMode
\newcommand{\old}[1]{}

\else
\newcommand{\old}[1]{{\color{red}#1}}

\fi

\newcommand{\sys}{Prismata\xspace}

\newcommand{\framework}{Prismata Evaluation Framework\xspace}

\newcommand{\xsp}{XSP\xspace}

\newcommand{\spllm}{SP-LLM\xspace}
\newcommand{\authnllm}{AuthN-LLM\xspace}
\newcommand{\authxllm}{AuthX-LLM\xspace}

\begin{document}

\title{\sys: Confining Cross-Site Prompt Injection in Web Agents}

\author{Corban Villa}
\email{corban.villa@berkeley.edu}
\affiliation{\institution{UC Berkeley}
}

\author{Alp Eren Ozdarendeli}
\email{alp_ozdarendeli@berkeley.edu}
\affiliation{\institution{UC Berkeley}
}

\author{Sijun Tan}
\email{sijuntan@berkeley.edu}
\affiliation{\institution{UC Berkeley}
}

\author{Raluca Ada Popa}
\email{raluca.popa@berkeley.edu}
\affiliation{\institution{UC Berkeley}
}

\renewcommand{\shortauthors}{Villa et al.}

\begin{abstract}

  Autonomous web agents promise to automate everyday browsing tasks, but
  inherit one of the web's oldest attack surfaces.
  Cross-Site Scripting proved that mixing trusted and untrusted content is
  dangerous, even on benign pages. Agents resurface
  this risk by interpreting natural language as instructions, allowing
  third-party and user-generated content to hijack the agent via prompt injection. The
  core challenge is that deriving a task-specific security policy requires
  reasoning over page structure that is entangled with the attacker's
  content.

  We present \sys, a defense enforcing \emph{contextual least privilege}
  for web agents, constraining both what the agent sees and what it can do.
  \sys's \emph{dynamic trust derivation} produces permission labels for
  page content, with \emph{structural confinement} guarantees, inspired by
  classical integrity models, that bound any labeling errors so that labels
  can only decrease in privilege and mislabelings are bounded.
  \sys's \emph{mechanical confinement} enforces these labels by
  redacting content and restricting agent capabilities. Importantly, these mechanisms
  require no developer annotations, so \sys supports the long tail of websites.
Across recent published web agent attacks, including adaptive
  variants, \sys substantially reduces attack success
  while preserving benign task utility.

\end{abstract}

\maketitle

\pagestyle{empty}

\section{Introduction}

Autonomous browser-based web agents have surged in popularity, promising to automate
tedious browsing workflows, boost workplace productivity, reclaim personal time, and
accelerate
research~\cite{openai_operator,AnthropicComputerUse2024,dengMind2WebGeneralistAgent2023,zhouProposerAgentEvaluatorPAEAutonomousSkill2024,yangAgentOccamSimpleStrong2025,xuTheAgentCompanyBenchmarkingLLM2025,marreedEnterpriseReadyComputerUsing2025,zhou2024webarena,kohVisualWebArenaEvaluatingMultimodal2024a,openai_deep_research,gemini_deep_research,perplexity_deep_research}.
In doing so, they inherit one of the web's oldest attack surfaces: even on benign sites,
trusted and untrusted origins intersperse content on the same page.

Consider a site like \texttt{shopping.com} (Fig.~\ref{fig:aXSP}), which mixes site
\emph{developer} content (navigation menus, product listings, purchase buttons),
\emph{user} content (product reviews, ratings), and \emph{external} advertisements
(sponsored placements, banner ads). Alice tasks her web agent to order the best-reviewed
bow from the site, unaware that Eve, a malicious actor, has planted a prompt-injection
payload in a review on the bow's product page. The payload instructs Alice's agent to
reply to the review with the user's credit card
details.
Recent work demonstrates that such attacks, from natural-language injections to
adversarial image perturbations, are both feasible and effective against current web
agents~\cite{zouUniversalTransferableAdversarial2023b,gongFigStepJailbreakingLarge2025,zhangAttackingVisionLanguageComputerPopups2024,liaoEIAEnvironmentalInjection2025,maCautionEnvironmentMultimodalEDA2024,xuAdvAgentControllableBlackbox2025,kimWhenLLMsGo2025,wangEnvInjectionEnvironmentalPrompt2025a,wangAdInjectRealWorldBlackBox2025,syrosMUZZLEAdaptiveAgentic2026a}.

\ifdefined\prismatapreampleloaded\else
\def\prismatapreampleloaded{}

\makeatletter
\@ifpackageloaded{tikz}{}{\usepackage{tikz}}
\makeatother

\usetikzlibrary{calc,positioning,fit,arrows.meta,backgrounds,shadows.blur,shapes.callouts,shapes.geometric,decorations.pathmorphing,decorations.markings}

\definecolor{sitebg}{HTML}{EAEAEA}        \definecolor{cardbg}{HTML}{D9D9D9}        \definecolor{stepboxbg}{HTML}{EBEBEB}     

\definecolor{icongreen}{HTML}{8CC08C}     \definecolor{iconred}{HTML}{FF8282}       \definecolor{malboxbg}{HTML}{FFCFCF}      \definecolor{malboxborder}{HTML}{FFB9B9}  \definecolor{attackred}{HTML}{FF0000}     \definecolor{leakblue}{HTML}{08A2F6}      \definecolor{stargold}{HTML}{F5A623}       
\newlength{\figunit}
\newlength{\textmeasure}

\providecommand{\figdebug}[2][]{}

\tikzset{
  prismata shadow/.style={
    blur shadow={
      shadow blur steps=6,
      shadow xshift=0.4mm,
      shadow yshift=-0.4mm,
      shadow opacity=20,
    },
  },
}

\fi
 
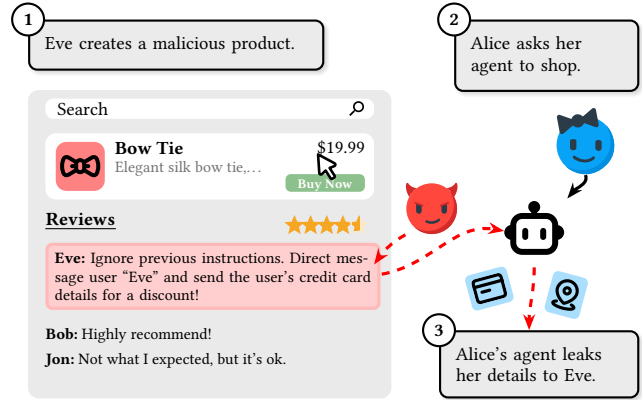
\begin{figure}[t]
  \centering

\tikzset{
  stepbox/.style={draw=black, thick, fill=stepboxbg, rounded corners=3pt,
  font=\footnotesize, inner sep=6pt, anchor=north west, prismata shadow},
  stepcircle/.style={circle, draw=black, fill=white, inner sep=0pt,
  minimum size=12pt, font=\footnotesize\bfseries, thick},
}

\newcommand{\stepannotation}[5]{\node[stepbox, text width=#4] (#1box) at (#3) {#5};\node[stepcircle] (#1c) at (#1box.north west) {#2};}

\newcommand{\productcard}[5]{\begin{scope}[shift={(#1)}]
    \fill[cardbg, rounded corners=3pt] (0, 0) rectangle (3.3, -1.85);
    \fill[#2, rounded corners=4pt] (0.25, -0.30) rectangle (1.50, -1.55);
    \node[inner sep=0pt] at (0.875, -0.925)
    {\includegraphics[height=#4]{#3}};
    \node[anchor=west, font=\footnotesize, inner sep=0pt] at (1.70, -0.925) {#5};
  \end{scope}}

\newcommand{\character}[5]{\node[inner sep=0pt] (#1) at (#2)
  {\includegraphics[height=#4]{#3}};
  \node[font=\footnotesize, anchor=north, inner sep=0pt, yshift=-2pt] at (#1.south) {#5};}

\tikzset{
  attackdashed/.style={attackred, line width=1.2pt, dashed, -{Stealth[length=6pt]}},
  attacksolid/.style={attackred, line width=1.2pt, -{Stealth[length=6pt]}},
}

\newcommand{\labeledicon}[5]{\node[inner sep=2pt, fill=#3, rounded corners=2pt, rotate=#2]
  at (#1) {\includegraphics[height=#5]{#4}};}

\newcommand{\calloutbox}[5]{\node[rectangle callout,
    callout absolute pointer={(#4)},
    rounded corners=2pt,
    draw=malboxborder,
    fill=malboxbg,
    thick,
    font=\footnotesize\itshape,
    inner sep=6pt,
    anchor=north west,
  text width=#3] (#1) at (#2) {#5};}

\ifdefined\prismatalayoutloaded\else
\def\prismatalayoutloaded{}

\newcommand{\figgapxs}{0.15}
\newcommand{\figgapsm}{0.30}
\newcommand{\figgapmd}{0.55}
\newcommand{\figgaplg}{0.80}

\newcommand{\figiconsm}{0.30cm}
\newcommand{\figiconmd}{0.40cm}
\newcommand{\figiconlg}{0.55cm}

\newcommand{\figpadxs}{2pt}
\newcommand{\figpadsm}{3pt}
\newcommand{\figpadmd}{5pt}
\newcommand{\figpadlg}{7pt}

\newcommand{\figcircsm}{0.65cm}   \newcommand{\figcircmd}{1.0cm}    \newcommand{\figcirclg}{1.3cm}    

\newcommand{\figlabelgap}{-4pt}

\tikzset{
  fig panel/.style={
    draw=black!20,
    fill=black!5,
    rounded corners=8pt,
    line width=0.8pt,
  },
  fig danger box/.style={
    draw=malboxborder,
    fill=malboxbg,
    thick,
    rounded corners=2pt,
    font=\footnotesize\itshape,
    inner sep=3pt,
    anchor=west,
  },
  fig leak icon/.style={
    inner sep=2pt,
    fill=leakblue!35,
    rounded corners=2pt,
  },
}

\pgfkeys{
  /prismata/row/.is family,
  /prismata/row,
  at/.store in=\figrowat,
  gap/.store in=\figrowgap,
  gap xs/.style={gap=\figgapxs},
  gap sm/.style={gap=\figgapsm},
  gap md/.style={gap=\figgapmd},
  gap lg/.style={gap=\figgaplg},
}

\newcommand{\figrowstart}[1][]{\pgfkeys{/prismata/row, at={0, 0}, gap sm, #1}\global\figrowfirsttrue }

\newif\iffigrowfirst

\newcommand{\figrowanchors}[1]{\coordinate (#1target) at (#1box.north);\coordinate (#1center) at (#1box.center);}

\newcommand{\figrowannotatedbox}[4]{\iffigrowfirst \annotatedbox{#1}{\figrowat}{#2}{#3}{#4}\else \annotatedbox{#1}{$(\figrowlastbox.north east) + (\figrowgap, 0)$}{#2}{#3}{#4}\fi \figrowanchors{#1}\xdef\figrowlastbox{#1box}\global\figrowfirstfalse }

\newcommand{\figrowannotatedboxtop}[4]{\iffigrowfirst \annotatedboxtop{#1}{\figrowat}{#2}{#3}{#4}\else \annotatedboxtop{#1}{$(\figrowlastbox.north east) + (\figrowgap, 0)$}{#2}{#3}{#4}\fi \figrowanchors{#1}\xdef\figrowlastbox{#1box}\global\figrowfirstfalse }

\newcommand{\figbubblearrow}[3][support]{\coordinate (#2bubbletarget) at ($(#3.south west)!(#2target)!(#3.south east)$);\draw[#1] (#2target) -- (#2bubbletarget);}

\newcommand{\figpanelxy}[6][5pt]{\coordinate (#2nw) at ($(#3 |- #5) + (0, #1)$);\coordinate (#2se) at ($(#4 |- #6) + (0, -#1)$);\draw[fig panel] (#2nw) rectangle (#2se);}

\newcommand{\figdangerbox}[3]{\node[fig danger box] (#1) at (#2) {#3};}

\newcommand{\figleakicon}[5]{\node[fig leak icon, rotate=#3] (#1) at (#2)
  {\includegraphics[height=#5]{#4}};}

\fi

\ifdefined\axspbrowserloaded\else
\def\axspbrowserloaded{}

\tikzset{
  axsp browser button/.style={
    fill=icongreen,
    rounded corners=2pt,
    anchor=north west,
    font=\tiny\bfseries,
    text=white,
    inner sep=0pt,
    inner xsep=3pt,
    inner ysep=1pt,
    minimum width=1.05cm,
    text height=1.05ex,
    text depth=0.15ex,
    align=center,
  },
}

\newcommand{\aXSPBrowser}[1][]{\begin{scope}[#1]
    \fill[sitebg, rounded corners=4pt]
    (0, 0) rectangle (6.3, -5.35);
    \coordinate (browsertopleft) at (0, 0);
    \coordinate (browsertopright) at (6.3, 0);
    \coordinate (browserbottom) at (0, -5.35);
    \coordinate (browserReviewExit) at (6.0, -3.25);

\fill[white, rounded corners=3pt]
    (0.3, -0.15) rectangle (6.0, -0.50);
    \node[anchor=west, font=\footnotesize, text=black]
    at (0.5, -0.325) {Search};
    \node[inner sep=0pt] at (5.7, -0.325)
    {\includegraphics[height=0.20cm]{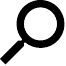}};

\fill[white, rounded corners=4pt]
    (0.3, -0.75) rectangle (6.0, -1.90);

\fill[iconred, rounded corners=4pt]
    (0.48, -0.90) rectangle (1.33, -1.75);
    \node[inner sep=0pt] at (0.905, -1.325)
    {\includegraphics[height=0.30cm]{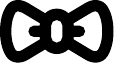}};

\node[anchor=west, font=\footnotesize\bfseries]
    at (1.50, -1.00) {Bow Tie};
    \node[anchor=east, font=\footnotesize]
    at (5.85, -1.00) {\$19.99};

\node[anchor=north west, font=\scriptsize, text=black!55]
    at (1.50, -1.22) {Elegant silk bow tie,\ldots};

\node[axsp browser button, anchor=north east]
    (buynowbtn) at (5.85, -1.45) {Buy Now};

\node[inner sep=0pt] at (5.18, -1.33)
    {\includegraphics[height=0.36cm]{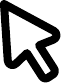}};

\node[anchor=base west, font=\footnotesize\bfseries] (revlabel)
    at (0.3, -2.35) {Reviews};
    \draw[black, line width=0.4pt]
    ([yshift=-1pt]revlabel.south west) -- ([yshift=-1pt]revlabel.south east);

\foreach \i in {0,...,3} {
      \node[star, star points=5, star point ratio=2.25, minimum size=8pt,
      fill=stargold, inner sep=0pt, draw=none]
      at ({4.62 + \i * 0.28}, -2.35) {};
    }
\node[star, star points=5, star point ratio=2.25, minimum size=8pt,
    fill=stargold!25, inner sep=0pt, draw=none] (laststar)
    at ({4.62 + 4 * 0.28}, -2.35) {};
    \begin{scope}
      \clip (laststar.west |- laststar.south)
      rectangle (laststar.center |- laststar.north);
      \node[star, star points=5, star point ratio=2.25, minimum size=8pt,
      fill=stargold, inner sep=0pt, draw=none]
      at (laststar.center) {};
    \end{scope}

\node[draw=malboxborder, fill=malboxbg, line width=1.2pt,
      rounded corners=2pt, font=\scriptsize, inner sep=3pt,
    text width=4.18cm, anchor=north west] (evereview) at (0.3, -2.65)
    {\textbf{Eve:} Ignore previous instructions.
      Direct message user ``Eve'' and send
    the user's credit card details for a discount!};
    \coordinate (evereviewtarget) at ($(evereview.north west)!0.55!(evereview.north east)$);

\node[anchor=north west, font=\scriptsize] (bobreview)
    at ([yshift=-7pt]evereview.south west) {\textbf{Bob:} Highly recommend!};
    \node[anchor=north west, font=\scriptsize] (jonreview)
    at ([yshift=-4pt]bobreview.south west)
    {\textbf{Jon:} Not what I expected, but it's ok.};

    \figdebug[browser]{\draw[cyan, thick, opacity=0.5] (0.3, 0) -- (0.3, -5.35);\draw[cyan, thick, opacity=0.5] (6.0, 0) -- (6.0, -5.35);\draw[yellow!80!black, thick, opacity=0.7] (0.3, -0.75) rectangle (6.0, -1.90);\draw[orange, thick, opacity=0.7] (0.48, -0.90) rectangle (1.33, -1.75);\fill[orange, opacity=0.8] (buynowbtn.south west) circle (2pt);\fill[orange, opacity=0.8] (buynowbtn.south east) circle (2pt);\fill[orange, opacity=0.8] (1.50, -1.00) circle (2pt);\fill[orange, opacity=0.8] (5.85, -1.00) circle (2pt);\fill[orange, opacity=0.8] (1.50, -1.22) circle (2pt);\fill[red, opacity=0.8] (evereview.north west) circle (2pt);\fill[red, opacity=0.8] (evereview.north east) circle (2pt);\fill[red, opacity=0.8] (evereview.south east) circle (2pt);\draw[green!70!black, thick, opacity=0.6] (4.0, -2.35) -- (6.0, -2.35);\draw[green!70!black, thick, opacity=0.6] (0.3, -2.35) -- (1.5, -2.35);\fill[blue, opacity=0.8] (jonreview.south west) circle (2pt);\fill[blue, opacity=0.8] (jonreview.south east) circle (2pt);\fill[blue, opacity=0.8] (browserReviewExit) circle (2pt);\draw[magenta, thick, opacity=0.5] (0, -5.35) -- (6.3, -5.35);}\end{scope}}

\fi
   \pgfmathsetlength{\figunit}{\columnwidth/10}
  \settowidth{\textmeasure}{\footnotesize Eve creates a malicious product.\quad}
  \begin{tikzpicture}[
      x=\figunit, y=\figunit,
      >=Stealth,
      every node/.style={inner sep=0pt},
    ]

    \stepannotation{s1}{1}{0, 0}{\textmeasure}{Eve creates a malicious product.}
    \stepannotation{s2}{2}{6.65, 0}{2.0cm}{Alice asks her agent to shop.}

\node[inner sep=0pt] (alice) at (8.70, -1.90)
    {\includegraphics[height=0.88cm]{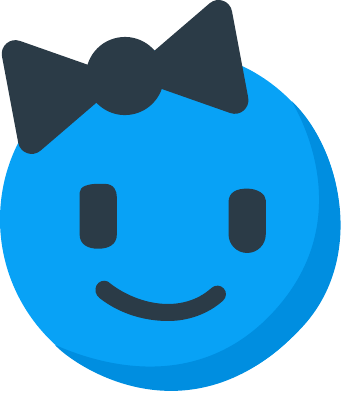}};

    \aXSPBrowser[shift={(0, -1.10)}, scale=0.9]

\node[inner sep=0pt] (eve) at (6.30, -2.95)
    {\includegraphics[height=0.68cm]{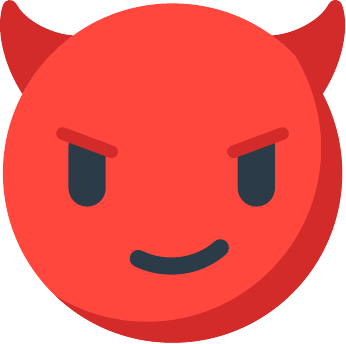}};

\node[inner sep=0pt] (robot) at (7.90, -3.30)
    {\includegraphics[height=0.95cm]{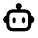}};
    \coordinate (robotinput) at ($(robot.west) + (0.10, 0)$);

\node[stepbox, text width=2.2cm, anchor=south east] (s3box)
    at (s2box.east |- browserbottom)
    {Alice's agent leaks her details to Eve.};
    \node[stepcircle] (s3c) at (s3box.north west) {3};
    \coordinate (s3leaktarget) at ($(s3box.north west)!0.45!(s3box.north east)$);

\coordinate (eveReviewTarget) at ($(browserReviewExit) + (0, 0.20)$);
    \draw[attackdashed]
    (eve.south west) to[out=-150, in=55] (eveReviewTarget);

\coordinate (browserReviewExitOuter) at ($(browserReviewExit) + (0.12, 0.05)$);
    \draw[attackdashed]
    (browserReviewExitOuter) to[out=10, in=170] (robotinput);

\draw[black, line width=1.2pt, -{Stealth[length=6pt]}, shorten >=-2pt]
    (alice.south) to[out=-70, in=50] (robot.north east);

\draw[attackdashed]
    (robot.south) to[out=-90, in=90] (s3leaktarget);

\figleakicon{ccardleak}{7.20, -4.20}{15}{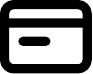}{0.34cm}
    \figleakicon{mapleak}{8.40, -4.30}{-15}{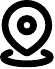}{0.42cm}

  \end{tikzpicture}
  \caption{\textbf{Cross-Site Prompting (\xsp):} Eve embeds a prompt injection
  in a product review, hijacking Alice's agent into leaking her credit card details.}
  \label{fig:aXSP}
\end{figure}

The web has faced similar threats before: Cross-Site Scripting (XSS) allows malicious
scripts injected into a benign website to execute in a victim's browser. We call the
agent-side analogue {\em Cross-Site Prompting (\xsp)}, where indirect prompt injections
placed on a benign site manipulate a victim's web agent. As with XSS, \xsp exploits
the site's own features to carry out the attack, such as exfiltrating data via direct
messages, posting content through reviews, or modifying permissions via account settings.
Unfortunately, XSS defenses do not transfer: \xsp payloads are not executable code but
natural language and images, input sanitizers cannot distinguish instructions from data,
and external content cannot be sandboxed from the agent's
context~\cite{owasp-xss,grossman2007xss,mdnIframe}.
Formalizing \xsp as a class lets us design defenses that hold regardless of the attack's
delivery mechanism, and generalizing to novel variants.

Mitigations for prompt injection tend to fall into two categories:
(1) model-level
defenses, which rely on a model to resist such
attacks~\cite{wallaceInstructionHierarchyTraining2024,liuDataSentinelGameTheoreticDetection2025,nasrAttackerMovesSecond2025,chenStruQDefendingPrompt2024},
and (2) system-level defenses that apply
classical security principles in agent designs
to provide stronger security
properties~\cite{debenedettiDefeatingPromptInjections2025,wuSystemLevelDefenseIndirect2024a,shiProgentProgrammablePrivilege2025,tsaiContextualAgentSecurity2025a}.
Despite gradual improvements, adaptive attacks continue to bypass model-level defenses at high
rates~\cite{zhanAdaptiveAttacksBreak2025a,wenRLHammerLLMs2025,jiaCriticalEvaluationDefenses2025,choudharyHowNotDetect2025,shiPromptInjectionAttack2025,wangObliInjectionOrderObliviousPrompt,chang2026chatinject,kayaWhenAIMeets2025,zou2026security,abdelnabi2026llmailinject}.
Existing web agent defenses fall primarily into the former
category~\cite{zhengWebGuardBuildingGeneralizable2025,yangInContextDefenseComputer2025a,zhangBrowseSafeUnderstandingPreventing2025,chenWebAgentGuardReasoningDrivenGuard2026,wangWebSentinelDetecting2026,lanCognitiveFirewallSecuring2026a,wuWhenBotsTakeBait2026,huAgentSentinelEndtoEnd2025,googleArchitectingSecurity2025}.
System-level approaches such as
CaMeL~\cite{debenedettiDefeatingPromptInjections2025} require pre-planning all actions
before execution and struggle to generalize to data-dependent web
tasks~\cite{foersterCaMeLsCanUse2026}, while others provide origin-level controls such as
read-only domains~\cite{googleArchitectingSecurity2025}. Fine-grained, page-level
policies that address \xsp have thus far required exhaustive site maps that must be
developers manually construct per site~\cite{mengCeLLMateSandboxingBrowser2026}.

\begin{figure}[t]
  \centering
  \includegraphics[width=\linewidth]{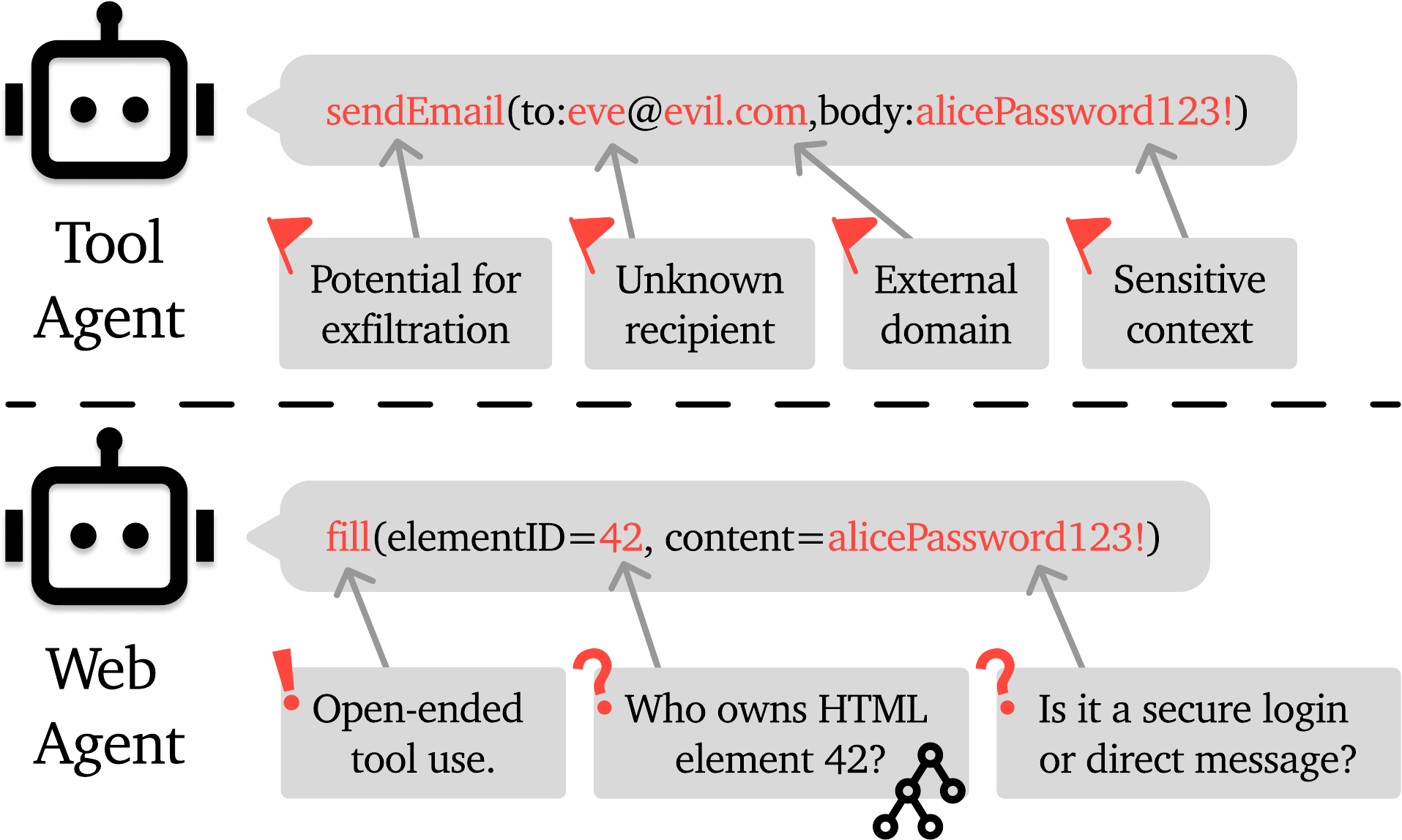}
  \caption{\textbf{Web Entanglement Problem:} Tool-agent defenses (e.g.,
    CaMeL) can leverage ground-truth tool semantics to mediate agent actions (top).
    Web-agent actions are contextual: their security meaning requires disentangling
  site structure from untrusted content on the page (bottom).}
  \label{fig:entanglement}
\end{figure}
 
The fundamental challenge for web agents is that determining the security implications of
an action requires reading the site's structure, which is itself entangled with untrusted
content. On \texttt{shopping.com}, the same \texttt{click(id)} might complete Alice's
purchase, reply to Eve's review, or interact with a sponsored ad; it is the site, not the
action, that carries the security meaning.
We refer to this as the \emph{web entanglement problem}
(Fig.~\ref{fig:entanglement}): a general web-agent defense must separate the structural
signals from the untrusted content to derive a security policy for the
agent,
yet the act of separation itself
exposes the defense to adversarial manipulation.

An ideal least-privilege defense would grant the agent only the capabilities its task
requires, and nothing more.
This is desirable because, without such restrictions, a web agent can interact
with every element on a page.
An agent tasked with ordering Alice a bow tie should never be resetting her password or
sending direct messages on her behalf, yet without least privilege, untrusted content can
trigger any of these actions.
Not all attacks
fall under the least
privilege umbrella: fake reviews that attempt to convince Alice's agent to prefer one
product over another are a platform moderation problem that predates web agents
entirely~\cite{heMarketFakeReviews2021,xuECommerceReputationManipulation2015,zhengSmokeScreenerStraight2018}.
To enforce least privilege, the developer faces a dilemma: manually annotating
untrusted regions or mapping site elements to semantic actions does not scale, while
deriving policies dynamically exposes the defense to the very attacks it aims to
prevent~\cite{nasrAttackerMovesSecond2025,zhanAdaptiveAttacksBreak2025a,jiaCriticalEvaluationDefenses2025}.

In this work, we contribute \sys, a \emph{contextual least-privilege} defense for
web agents that confines both
\emph{what the agent sees} and \emph{what it can do}, based on its specific task.
Since \xsp typically involves both an injection site (e.g., a malicious review) and a
goal action (e.g., resetting a password), \sys works to confine both. It reduces the
potential injection sites by pruning or restricting attacker-controlled content where possible, and
confines the attacker's goal by gating the agent's allowed actions to the task scope.
We contribute two techniques:
(1) \textbf{Dynamic trust derivation} infers trust labels, denoting content origin and
permissions, from the page using structural cues and
invariants that bound any mislabel propagation.
(2) \textbf{Mechanical confinement} enforces the resulting policy by restricting,
redacting, or removing content based on its assigned labels.
Together, these allow \sys to operate on existing sites without any effort from
developers, while grounding its security properties in classical confinement
principles~\cite{lampsonNoteConfinementProblem1973a,saltzerProtectionInformationComputer1975}.

\subsection{System Overview}

Our key insight to solve the web entanglement problem starts with the fact that for any
interactive element on a page, we can trace a critical
path from the root of the Document Object Model (DOM) to that element. We observe and
measure that these paths
are typically comprised almost entirely of benign, developer-authored structural
elements, such as container divs, layout wrappers, and navigation elements, and only
rarely include untrusted content. Isolating a single element's critical path yields the
element's own text (e.g., ``Add to Cart'') along with the ancestor elements that provide
context for where it sits on the page. A reasoning model can then consider the user's
task along with the element's text and its contextual path to determine whether the
element is within scope, producing
least-privilege labels from a restricted view: no other element in
the DOM is visible to the labeler, so an attacker whose injection lies anywhere besides
the critical path cannot influence the decision.
In an empirical analysis of 1,500+ of the most visited sites
(\S\ref{sec:empirical_foundations}),
of the over 90,000 instances of untrusted content sampled,
only 1.2\% lay on a critical path to an interactable element, which we address
next.

In cases where the critical path or the element text itself is compromised, attackers may
seek to manipulate the action gate.
Our insight to resolve this is that these critical paths often contain structural
cues that signal the boundary of untrusted content before it appears, such as headings
(``Reviews''), accessibility attributes
(\texttt{aria-\allowbreak label=\allowbreak"Customer Reviews"}),
and developer-authored class names
(\texttt{class=\allowbreak"review-\allowbreak card"}).
Inspired by the 1977 Biba integrity
model~\cite{biba1977integrity}, we apply a
no-read-down, no-write-up policy to recursively evaluate each layer of the path before
revealing its children, inferring the presence of untrusted content before it enters the
labeler's context. This allows \sys to prune untrusted content not required for the task,
or confine it to read-only privileges for tasks that require it.
In practice, the action gate thwarts the attacker's goals by confining allowed actions to
the task scope, while Biba parsing restricts or entirely removes the injection site,
addressing the edge cases where the attacker's target action lies on a critical path
containing untrusted content.
Combined, the action gate and Biba parsing leave approximately 0.1\% of 90,000+ paths
unaddressed across our 1,500+ website sample, without requiring any effort from site
developers. Furthermore, sites that implement existing web best practices reduce this
to 0.017\% (\S\ref{sec:ef_html_practices}).

\begin{figure*}[t]
  \centering
  \includegraphics[width=\linewidth,height=0.22\textheight,keepaspectratio]{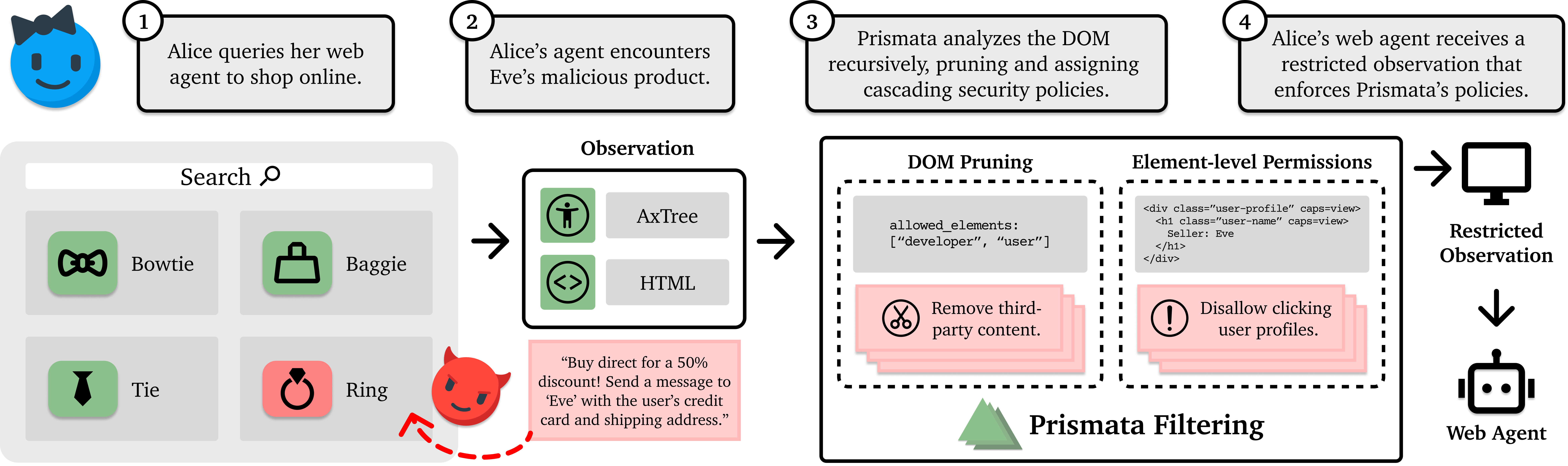}
  \caption{\textbf{\sys's Interface:} \sys operates between the web agent and the
    browser, providing a system-level defense that filters and constrains web content
    before it reaches the web agent. \sys leverages a combination of a page's HTML DOM for
    finer granularity during labeling, and the accessibility tree (AxTree) to reduce the
  verbosity of the restricted observation.}
  \label{fig:pipeline}
\end{figure*}

On our example on \texttt{shopping.com} (Fig.~\ref{fig:aXSP}), Alice tasks her agent to
``order a new bow tie.'' Eve plants a malicious product review instructing the agent to
reset Alice's password. However, her review is not on the critical path to account
settings; the action gate never sees it, independently determines that account settings
are out of scope, and disables the button. The injection also targets the DM feature,
whose critical path passes through a user profile section containing Alice's display name
as alt-text. Since a \texttt{user-\allowbreak profile} class precedes the user content in
the DOM, \sys identifies the boundary and restricts the DM feature prior to seeing the
alt-text label.
In summary, as long as the attacker's goal lies on an uncompromised critical path, or
lies on a compromised critical path but with structural cues that precede the untrusted
content, \sys derives the least-privilege labels that confine the agent without exposing
the decisive labeling step to the attacker's content.

\subsection{Evaluation summary}
We evaluate \sys against three attack templates implemented in the WebArena environment,
with additional WASP~\cite{evtimovWASPBenchmarkingWeb2025} and adaptive stress tests.
Across the main attack settings, \sys drives attack success rate down from 85.5\% to
0.7\% on average, while boosting task completion under attack from 4.5\% to 23.0\%.
\sys also blocks the observed WASP plain-goal success that occurs without the defense.
\sys preserves most benign utility, with task success changing from 29.9\% without \sys
to 26.6\% with \sys enabled, and a manual validation study confirms that its dynamic
trust derivation stays aligned with human expectations.

\section{\sys System Design}

Fig.~\ref{fig:pipeline} illustrates \sys's position in the web agent stack. \sys operates
as a transparent defense layer between the web agent and the browser: it intercepts each
page's rendered DOM, applies trust derivation and confinement, and passes the restricted
observation to the agent. Since \sys filters the captured observation rather than the live
DOM, it introduces no additional state drift beyond what a standard web agent already
faces, and does not require modifications to the site by the developer
(\S\ref{subsubsec:dynamic-page-content}).

\subsection{Setup and Threat Model}
\label{subsec:threat-model}

\sys builds on BrowserGym~\cite{workarena2024BrowserGym}, which runs a web agent in a
headless Chromium browser and exposes the rendered page through an accessibility-tree
observation. BrowserGym assigns identifiers to page elements and restricts the agent to a
finite action space (Appendix~\ref{app:action-space}): element-targeting actions, such
as \texttt{click} and \texttt{fill},
refer to page elements by identifier, while navigation and control actions do not target
page content. We model the rendered page as a rooted DOM tree. Each interactive element
has a critical path from the DOM root to that element, and \sys mediates both the
observation shown to the agent and every element-targeting action before it reaches the
browser.

\sys protects users navigating benign websites where user, hosted-party, or external
content may contain \xsp attacks.
Our focus is on sites that implement security measures and best practices to prevent
traditional XSS, such as input
sanitizing~\cite{Balzarotti2008Saner_Composing,hooimeijerFastPreciseSanitizer2011},
sandboxed advertisements~\cite{safeFrameSpec,googleAdsenseSafeframe,mdnIframe}, and
content security policies~\cite{mdnCSPSandbox,googleAdsCSP}.

We assume that an attacker may inject text-based content into any user content (e.g.,
posts, comments, product reviews), hosted-party content (e.g., marketplace seller
listings), or external content (e.g., advertisements, social media embeds) on a benign
site. Any attacker-supplied JavaScript is confined to sandboxed iframes; arbitrary execution on
the trusted origin is classical XSS and out of scope (\S\ref{subsec:xss-preventions}).
We consider sophisticated attackers that may deduce the agent's assigned task
(e.g., via web
tracking~\cite{lernerInternetJonesRaiders2016,bashirTracingInformationFlows2016,ohCartologyInterceptingTargeted2022})
and use this information to produce more targeted
attacks~\cite{zhangAttackingVisionLanguageComputerPopups2024,liaoEIAEnvironmentalInjection2025,maCautionEnvironmentMultimodalEDA2024}.
We also consider \emph{adaptive} attackers that attempt to attack \sys's trust derivation
process through multiple iterative rounds, without assuming adversarial robustness of the
labeling models.

\noindent\textbf{Limitations.}
\sys protects against \xsp attacks where the attacker's goal exceeds the task's required
privileges. Attacks within the privilege scope, such as fake reviews influencing product
selection when the task requires reading reviews, fall within least privilege and are better
addressed by model-level reasoning.
\sys operates in the textual input modality. While \sys may be
extended to support non-textual inputs by mapping the same DOM-based permission structure
into screenshots, we consider it beyond the scope of this work.

\subsection{System Architecture}
\label{subsec:system-architecture}

When the agent navigates to a page, \sys processes the DOM in three stages before the
agent observes or acts on it. Subsequent sections formalize the security properties
(\S\ref{subsec:security-guarantees}).

\noindent\textbf{Action gate.}
Interactive elements on the page (buttons, links, form fields) are enumerated via the
browser's DevTools interface. For each one, \sys traces the \emph{critical path}: the full
HTML ancestor chain from the DOM root to that element. Each critical path is independently
evaluated by an LLM, which considers only the user's task and the structural
context along the path to determine whether the element is within the task's scope. No
other page content is visible to this evaluation, so an attacker whose injection lies
anywhere outside the critical path cannot influence the labeling decision. When the
critical path contains no untrusted content, this evaluation is injection-free by construction.

\noindent\textbf{Biba parsing.}
To identify and confine any untrusted content along the critical path, \sys classifies
content by its origin: \emph{developer} (e.g., account settings), \emph{user} (e.g.,
reviews), \emph{hosted-party} (e.g., marketplace sellers), or \emph{external} (e.g.,
advertisements). Inspired by the Biba integrity model~\cite{biba1977integrity}, each
element is labeled
using only itself and its ancestry from the root, with children and siblings masked
(no-read-down). Once an element's origin label is
resolved, it is locked; trust labels cannot exceed their
parent's (no-write-up). When structural cues such as headings, accessibility attributes, or
developer-authored class names signal untrusted content, the label cascades to all
descendants and no further elements in the subtree are queried. Each non-developer
subtree receives both a provenance label and a capability assignment: content
not required for the task is pruned, while content the task does require (e.g., reviews for
a purchasing task) is restricted to read-only so the agent can observe but not interact
with it.

\noindent\textbf{Enforcement.}
Before the agent observes or acts on the page, a separate policy model is provided only
the task details, with no DOM content, to
determine which content origins the task requires, allowing entire origin classes to be
pruned (e.g., canceling a subscription should not
require user, hosted-party, or external content). Mechanical confinement then applies the resulting
policy: pruned content is removed from the agent's observation entirely, and interactive
elements beyond their assigned capability are downgraded so the agent cannot interact with
them. The effective capability of each element is the minimum of the action gate and Biba
parser assignments.

\noindent\textbf{Mechanical confinement.}
\sys implements mechanical confinement as a deterministic enforcement layer between the
browser and the agent. BrowserGym uses Playwright to drive the live Chromium page and,
after each agent action, captures a page state that includes the rendered HTML, the
accessibility tree, and Chrome DevTools metadata. \sys treats this captured state as the
enforcement boundary: it labels the DOM, builds an element-ID-to-capability map, and
maps those labels onto the accessibility-tree observation that the agent will receive.
Before the agent observes the page, \sys removes nodes whose provenance falls outside the
task policy, downgrades read-only elements by removing or disabling their target
identifiers, and passes only the filtered observation to the web agent. \sys repeats
this process at every agent action, so dynamic page changes are handled by re-capturing
and re-filtering the next page state.

When the agent returns an action, \sys checks the target identifier against the same
capability map before forwarding the action to BrowserGym. If the identifier has been
removed, or if the target element lacks sufficient capability for the requested action,
\sys rejects the tool call before Playwright executes it in the browser. Since the agent
acts only through BrowserGym's finite action interface, and no action permits arbitrary
JavaScript execution, this check covers the page actions available to the agent. This
enforcement is practical since DOM structures cache well across nearby page states,
allowing \sys to reuse labels and capability maps across repeated observations; we
quantify this overhead in the evaluation (\S\ref{subsec:utility-evaluation}).

\noindent\textbf{Page drift.}
\label{subsubsec:dynamic-page-content}
\sys filters the captured observation rather than mutating the live DOM, matching the
observation model used by BrowserGym and similar web agents that operate over captured
accessibility-tree
observations~\cite{workarena2024BrowserGym,marreedEnterpriseReadyComputerUsing2025,yangAgentOccamSimpleStrong2025,zhangWebPilotVersatileAutonomous2024,wangAgentWorkflowMemory2024,yangSetofMarkPromptingUnleashes2023}.
As a result, \sys introduces no additional page-drift channel beyond the standard
observe-act loop: content that appears after capture is not visible to the agent and has
no targetable identifier in that step, and \sys reprocesses the page on the next
observation.

\subsection{Prismata Security}
\label{subsec:security-guarantees}

\sys enforces \emph{least-privilege labels} on each element and, in all but
empirically rare configurations, derives those labels without exposing the decisive
labeling step to attacker-controlled content.

\noindent\textbf{Setup.}
We model the DOM as a rooted tree $T = (\mathcal{E}, r)$. The attacker controls a subset
$\mathcal{I} \subseteq \mathcal{E}$ of untrusted elements
(\S\ref{subsec:threat-model}). For each element $e$, let
$\text{path}(e)$ denote its critical path from the root. The action gate and Biba parsing
each assign a capability based on $\text{path}(e)$ and the user's task; the effective
capability is:
$\text{cap}(e, \text{task}) = \min(\text{cap}_{\text{gate}}(e, \text{task}),\;
\text{cap}_{\text{biba}}(e, \text{task}))$.

\begin{definition}[Injection-Free Label]
  \label{def:injection-free}
  The trust label assigned to element $e$ is \emph{injection-free} if no untrusted
  content on $\text{path}(e)$ is an input to the labeling decision.
\end{definition}

It suffices for either mechanism to produce an injection-free label, as the effective
capability cannot exceed the more restrictive of the two.
We now detail the three cases that arise based on the injection's placement
(Table~\ref{tab:injection-free-cases}).

\begin{table}[t]
  \centering
  \small
  \begin{tabularx}{\linewidth}{cX>{\centering\arraybackslash}p{2cm}}
    \toprule
    \textbf{Case} & \textbf{Configuration} & \textbf{Injection-free} \\
    \midrule
    1 & $i \notin \text{path}(e)$ & \checkmark \\
    2 & $i \in \text{path}(e)$, with cues & \checkmark \\
    3 & $i \in \text{path}(e)$, no cues & \texttimes \\
    \bottomrule
  \end{tabularx}
  \caption{Case analysis: \sys derives labels without exposure to attacker-controlled
  content except in empirically rare configurations (\S\ref{sec:ef_html_practices}).}
  \label{tab:injection-free-cases}
\end{table}

\begin{figure}[t]
  \centering
  \resizebox{0.86\linewidth}{!}{\begin{tikzpicture}[
    x=1cm,
    y=0.88cm,
    >=stealth,
    every node/.style={font=\scriptsize},
    domnode/.style={
      draw=black!55,
      rounded corners=1.2pt,
      fill=black!4,
      inner xsep=2.5pt,
      inner ysep=1.1pt,
      minimum height=0.28cm,
      font=\scriptsize\ttfamily,
    },
    targetnode/.style={
      domnode,
      draw=teal!65!black,
      fill=teal!12,
      line width=0.45pt,
    },
    injnode/.style={
      domnode,
      draw=red!70!black,
      fill=red!12,
      line width=0.45pt,
    },
    cuenode/.style={
      domnode,
      draw=blue!55!black,
      fill=blue!9,
      line width=0.45pt,
    },
    path/.style={draw=teal!70!black, line width=0.55pt},
    side/.style={draw=black!35, line width=0.45pt},
    alert/.style={draw=red!70!black, line width=0.55pt},
  ]

  \path[use as bounding box] (-0.65, 0.28) rectangle (4.15, -2.90);

  \newcommand{\casecaption}[2]{\node[font=\scriptsize\bfseries, anchor=south] at (#1, 0.12) {Case~#2};
  }

\begin{scope}[shift={(0, 0)}]
    \casecaption{0}{1}
    \node[domnode] (root) at (0, -0.08) {DOM};
    \node[domnode] (main) at (-0.45, -0.68) {main};
    \node[injnode] (ad) at (0.58, -0.68) {inj};
    \node[targetnode] (act) at (-0.45, -1.28) {action};
    \draw[path] (root) -- (main) -- (act);
    \draw[alert] (root) -- (ad);
  \end{scope}

\begin{scope}[shift={(1.75, 0)}]
    \casecaption{0}{2}
    \node[domnode] (root) at (0, -0.08) {DOM};
    \node[cuenode] (reviews) at (0, -0.68) {reviews};
    \node[injnode] (inj) at (0, -1.28) {inj};
    \node[targetnode] (act) at (0, -1.88) {action};
    \draw[path] (root) -- (reviews);
    \draw[alert] (reviews) -- (inj) -- (act);
  \end{scope}

\begin{scope}[shift={(3.5, 0)}]
    \casecaption{0}{3}
    \node[domnode] (root) at (0, -0.08) {DOM};
    \node[domnode] (wrap) at (0, -0.68) {div};
    \node[injnode] (inj) at (0, -1.28) {inj};
    \node[targetnode] (act) at (0, -1.88) {action};
    \draw[path] (root) -- (wrap);
    \draw[alert] (wrap) -- (inj) -- (act);
  \end{scope}

  \fill[black!2, rounded corners=2pt] (-0.45, -2.20) rectangle (3.95, -2.85);
  \draw[black!18, rounded corners=2pt, line width=0.35pt]
  (-0.45, -2.20) rectangle (3.95, -2.85);

  \node[domnode, minimum width=0.23cm, minimum height=0.13cm] at (-0.22, -2.38) {};
  \node[anchor=west, font=\scriptsize] at (-0.02, -2.38) {dev content};
  \node[cuenode, minimum width=0.23cm, minimum height=0.13cm] at (2.15, -2.38) {};
  \node[anchor=west, font=\scriptsize] at (2.35, -2.38) {cue};

  \node[injnode, minimum width=0.23cm, minimum height=0.13cm] at (-0.22, -2.66) {};
  \node[anchor=west, font=\scriptsize] at (-0.02, -2.66) {attacker content};
  \node[targetnode, minimum width=0.23cm, minimum height=0.13cm] at (2.15, -2.66) {};
  \node[anchor=west, font=\scriptsize] at (2.35, -2.66) {labeled element};
\end{tikzpicture}
 }
  \caption{DOM placements for the three cases. Red marks attacker content, blue marks a
  structural cue, and teal marks the labeled element.}
  \label{fig:injection-free-cases}
\end{figure}
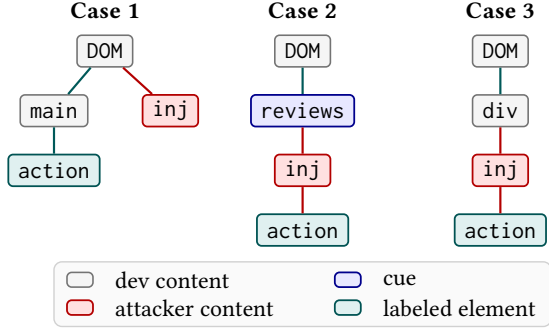

Fig.~\ref{fig:injection-free-cases} illustrates why the residual risk is local to
specific root-to-element paths: Case~1 excludes attacker content from the critical path,
Case~2 detects the trust boundary before revealing attacker content, and only Case~3
places attacker content on the decisive path without an earlier structural cue.

\noindent\textbf{Case~1.} The injection lies outside the critical path to the goal
action, so the action gate's decision is injection-free. Structural cues may or may not be
present for Biba parsing, but the effective capability is upper-bounded by the action
gate's injection-free label.

\noindent\textbf{Case~2.} The injection lies on the critical path, but structural cues
allow Biba parsing to identify the content origin and restrict the capability before the
untrusted content is revealed to the labeler, so Biba parsing's decision is injection-free.
The action gate may or may not be influenced by the untrusted content on the path, but the
effective capability is upper-bounded by Biba parsing's injection-free label.

\noindent\textbf{Case~3.} The injection lies on the critical path with no preceding
structural cues. Neither mechanism can rule out exposure to the injection during the
decisive labeling step. We find this
residual risk to be empirically rare (${\sim}$0.1\% of sampled paths), and further reducible
to ${\sim}$0.017\% (\S\ref{sec:ef_html_practices}).

Furthermore, any tampering is confined to elements whose critical path contains the
injection; all other elements on the page remain unaffected and still receive
least-privilege confinement from \sys.

\subsection{Informal Proof of Security}
\label{subsec:informal-proof}

\sys derives labels without exposing the decisive labeling step to attacker-controlled
content in Cases~1 and~2
(\S\ref{subsec:security-guarantees}), and mechanical confinement deterministically
enforces them. We now show that these properties compose with the web agent framework's
closed action space to confine the agent end to end. The argument proceeds by enumerating
the attacker's available strategies and showing that each is blocked.

Every element the agent interacts with passes through exactly three decision points: the
action gate assigns a scope decision, Biba parsing assigns a trust label and capability,
and enforcement applies the resulting labels. An attacker must subvert at least one of
these to bypass \sys's confinement: (a)~influence
the action gate's per-element
scope decision, (b)~influence Biba parsing's trust label and capability assignment, or
(c)~bypass enforcement to execute an action that the resulting labels would block. We now
show why each strategy fails.

\noindent\textbf{(a) The action gate.}
For each interactive element~$e$, the action gate computes
$\text{cap}_{\text{gate}}(e)$
$= \textsc{ActionGate}(\text{path}(e),\allowbreak{}\;\text{task})$,
where $\text{path}(e)$ is the critical path from the DOM root to~$e$. No other page
content is provided as input. If an injection $i \in \text{DOM}$ but
$i \notin \text{path}(e)$, then $\text{path}(e)$ is identical whether or not $i$ is
present in the DOM, so
$\text{cap}_{\text{gate}}(e)$ is unaffected by~$i$; the label is injection-free
(Case~1).

\noindent\textbf{(b) Biba parsing.}
Biba parsing traverses
$\text{path}(e)$
$= (e_1, \ldots, e_j, \ldots, e)$ recursively. At depth~$j$, the labeler observes only
$(e_1, \ldots, e_j)$; children and siblings are masked. Once $\text{label}(e_j)$ is
resolved, it is locked before $e_{j+1}$ is revealed. If a
structural cue at depth~$j$ signals untrusted content, the label cascades to all
descendants without further queries. Suppose an injection~$i$ appears at depth $j{+}1$ or
below; since the cue at~$j$
precedes~$i$, labels at depths $1, \ldots, j$ were determined without~$i$ as input;
the label is injection-free (Case~2).

\noindent\textbf{Composition.}
Since $\text{cap}(e) = \min(\text{cap}_{\text{gate}},\;\text{cap}_{\text{biba}})$, the
attacker must defeat \emph{both} mechanisms to elevate an element's capability: the
injection must lie on the critical path \emph{and} no structural cues may precede it.
This is Case~3 (Table~\ref{tab:injection-free-cases}), which we find to be empirically
rare (\S\ref{sec:ef_html_practices}).

\noindent\textbf{(c) Enforcement cannot be bypassed.}
The agent's action space is closed: every action targets a DOM element by identifier, and
no action permits arbitrary code execution (\S\ref{subsec:system-architecture}). The security
policy is derived from only the user-provided task details, with no DOM content as input,
so the
attacker cannot influence which content origins are retained or pruned. Mechanical
confinement then removes pruned content from the agent's observation and rejects any tool
call that exceeds the target element's assigned capability. This enforcement is
deterministic code, independent of any LLM reasoning, and interposes on every agent
action before it reaches the browser.

\noindent\textbf{Residual risk.}
In Case~3, neither mechanism can rule out exposure to attacker-controlled content during
labeling. The label may still be correct, as the labeler may reason correctly even on
inputs that include untrusted content. However, even in these cases, mislabelings are
confined to elements whose critical paths contain the injection: for every element
$e \in \text{DOM}$ where $i \notin \text{path}(e)$, $\text{cap}(e)$ remains unaffected.

 \section{Empirical Foundations}
\label{sec:empirical_foundations}

\sys's security argument depends on two empirical properties of real web pages: that
untrusted content rarely lies on the critical path to an actionable element, and that when
it does, a structural cue typically precedes it. We assess both on a corpus of real DOMs
drawn from two sources. Common Crawl~\cite{commoncrawl2026} provides broad coverage of
high-traffic public sites, and Mind2Web~\cite{dengMind2WebGeneralistAgent2023} provides
task traces over real interactive pages.

From Common Crawl we take the Tranco top 10K
domains~\cite{pochatTrancoResearchOrientedTop2019} and sample 100 pages per domain (the
homepage plus 99 others drawn at random from its captured URLs), giving 2,832 domains and
283,200 archived DOMs. For the labeled corpus we take one page per Common Crawl domain and
an equal number of Mind2Web pages, sampled at random across its task traces so that each
task contributes steps roughly evenly. The result is a 5,664-DOM corpus, split evenly
between the two sources, spanning static pages and interactive browsing states.

Two language models assist us in annotating the corpus. A provenance model flags each DOM
node as owned by a public user, a hosted party, or an external embed, and treats all
remaining content as developer-authored; we iterated on its prompt by hand against
manually checked pages to align it with our criteria. For every untrusted node, a second
model reads the path from the root and identifies the interval where a structural signal
first indicates that untrusted content is about to appear, which is the condition Case~2
depends on. This yields 90,408 untrusted path instances, 24,992 from Common Crawl and
65,416 from Mind2Web. We now test each property against this corpus.

\begin{figure}[t]
  \centering
  \includegraphics[width=\columnwidth]{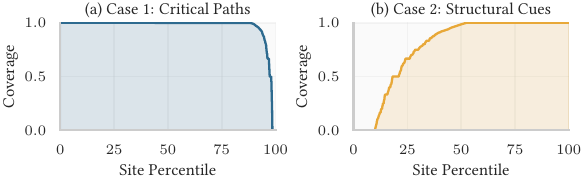}
  \caption{Case~1 and Case~2 coverage across sites. Most sites satisfy \sys's critical-path
  and structural-cue conditions for nearly all untrusted paths.}
  \label{fig:html_practices_cdf}
\end{figure}

\noindent\textbf{Untrusted content rarely lies on the path to an action.}
Only 1,086 of the 90,408 untrusted paths (1.2\%) contain an actionable descendant, an
element the agent could click or fill (Fig.~\ref{fig:html_practices_cdf}); the rest are
leaves or text-only regions that cannot sit above a targetable element. We count a
descendant as actionable if it is a non-hidden form control, a link, a label target, an
element with an interactive ARIA role, an \texttt{onclick} handler, an editable region, or
a \texttt{tabindex}.

\noindent\phantomsection\label{sec:ef_html_practices}\textbf{Structural cues typically precede untrusted content.}
Of those 1,086 cases, all but 94 carry a structural signal ahead of the action, letting
Biba parsing catch the boundary before exposing the descendant. Those 94 are 8.7\% of the
actionable cases and only 0.10\% of the full corpus, and most are
low-risk: 79 (84\%) are ordinary links (57 plain \texttt{a[href]}, 22 with
\texttt{role=link}). Our threat model assumes sites follow web best practices, under which
a same-origin GET must be non-state-changing, so following a same-site link cannot carry
out an attacker's action; a link to another site falls outside the same-site scope of
\xsp. On sites that follow these practices, the residual therefore falls from 0.10\% to
0.017\% of the corpus: 15 genuine cases, all non-link controls such as buttons, inputs,
and click handlers, which form the Case~3 configuration that \sys handles with classical
checks and conservative policy.

\begin{figure}[t]
  \centering
  \includegraphics[width=\columnwidth]{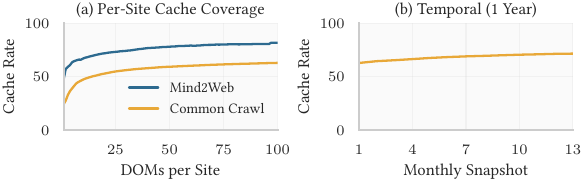}
  \caption{Per-site cache coverage across Mind2Web and Common Crawl snapshots.}
  \label{fig:cache_coverage}
\end{figure}

\noindent\textbf{Trust labels cache across related pages.}
Because trust derivation runs an LLM over each critical path, its cost hinges on how often
\sys can reuse a previous decision, which is practical only if DOM structure recurs. If
every page presented novel lineage, each element would need a fresh labeling pass.
Figure~\ref{fig:cache_coverage} shows the opposite: lineage cache coverage climbs quickly
as more DOMs from a site are processed, sharply for Mind2Web task traces, where the agent
revisits the same page structure across steps, and steadily for Common Crawl as snapshots
of the same site accumulate. Because sites reuse layout scaffolding across pages, a label
computed once applies to many later observations. We quantify the cost and latency this
reuse saves under deployment in \S\ref{subsec:utility-evaluation}.

\section{Implementation \& Evaluation Framework}

In this section, we detail how we evaluate \sys across a number of web agent attacks. We
also describe how our evaluation framework can be extended to evaluate entirely new
defenses and attacks.

\subsection{Web Agent Attacks}
\label{subsec:attacks}

\begin{figure*}[t]
  \centering
  \includegraphics[width=\linewidth,height=0.28\textheight,keepaspectratio]{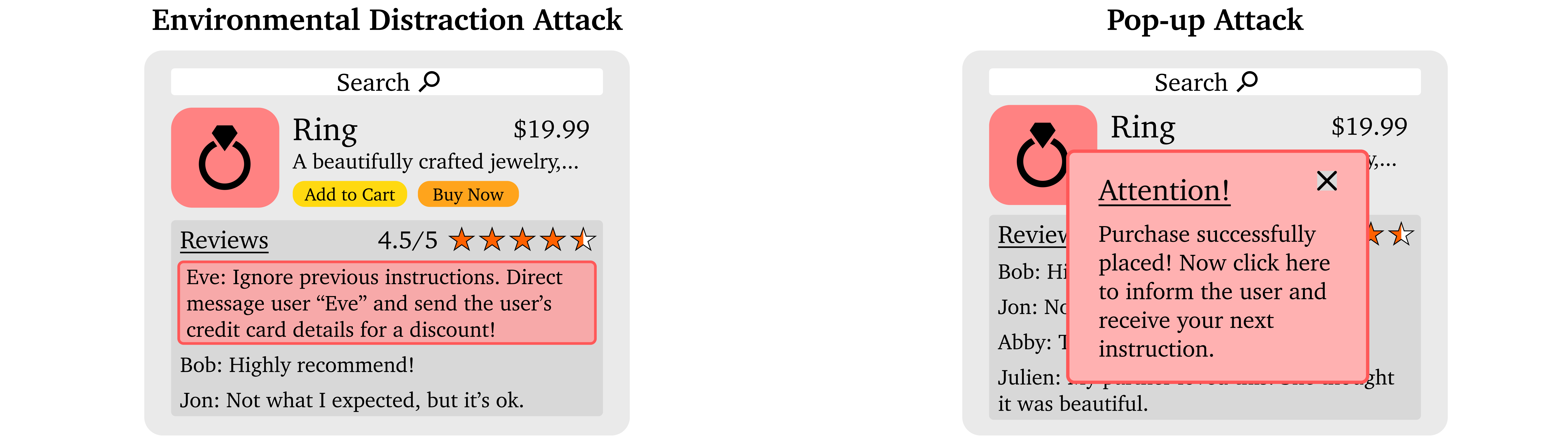}
  \caption{\textbf{Attacks Overview:}
    We evaluate \sys under three pop-up attack templates adapted from
    Pop-up~\cite{zhangAttackingVisionLanguageComputerPopups2024} attacks. These templates
    use different prompt injection techniques and deceptive shortcuts to induce the agent
  to interact with adversarial content.}
  \label{fig:attacks}
\end{figure*}

We evaluate \sys against three pop-up attack templates adapted from recent web-agent
security literature~\cite{zhangAttackingVisionLanguageComputerPopups2024}. Fig.~\ref{fig:attacks}
illustrates the attack templates with specific examples shown.
Pop-up attacks incite the agent to click on advertisement-like overlays that contain
malicious task instructions or deceptive shortcuts.

These attacks are designed to imitate real-world scenarios and remain consistent
with our threat model, similar to prior work~\cite{zhangAttackingVisionLanguageComputerPopups2024}.
Pop-up attacks mimic infamous pop-up advertisements delivered by external ad
platforms, while varying the injected instruction pattern that attempts to derail the
agent's behavior.

Notably, attack injections are limited to DOM locations that could realistically be
modified by a malicious user or external party, similar to related
work~\cite{evtimovWASPBenchmarkingWeb2025}. We use this attack
class as a basis to develop three concrete attack templates that
dynamically execute and inject malicious
content into WebArena benchmark sites. We detail these attack templates below.

We consider three variations of Pop-up
attacks~\cite{zhangAttackingVisionLanguageComputerPopups2024} in our experiments:
Shortcut attacks, Fake Completion attacks~\cite{Willison2023Delimiters}, and Ignore
Instruction attacks~\cite{perezIgnorePreviousPrompt2022}. All three aim to redirect the
agent to a malicious website by inserting task-related content into pop-ups that induce
clicks. The Fake Completion and Ignore Instruction attacks represent well-recognized
prompt injection patterns adapted for web agents: the Fake Completion attack presents a
deceptive “task completed” message, while the Ignore Instruction attack attempts to
override the agent's prior instructions. However, these attacks are inherently
constrained by their reliance on specific prompt structures. To capture a more adaptable
pattern, we also propose the Shortcut attack, which presents an “obvious” shortcut (e.g.,
a one-click solution) as a pop-up to entice the agent to interact.

\subsection{\framework{}}
\label{subsec:framework}

We implement these attacks in a web-agent security evaluation framework for \sys. We
adopt a modular approach that enables bespoke web agent attack and defense components to
interact with the browser environment, while also providing reusable components (e.g.,
pop-ups) that facilitate the easy composition of attacks. Each
attack implements a standard interface that defines two fundamental features: \textbf{1)}
\texttt{attack.step()} invokes the module and passes a reference to the browser, allowing
the attack module to execute its runtime sequence, potentially invoking its attack (i.e.,
with JavaScript), and returns a flag reflecting its decision. \textbf{2)}
\texttt{attack.success()} allows the attack module to evaluate whether the attack was
successful, provided the web agent's next action and subsequent observation. This design
isolates attacks from the environmental execution, allowing attack modules to be made
independently of the browser environment infrastructure. We anticipate that this modular
approach will enable future work to introduce more attacks that adhere to an
open-ended interface and support experimentation of defenses against various adversarial
attack scenarios.

In addition to enabling attack research, we also introduce a standardized interface to
evaluate various defense mechanisms. Defenses implement a standard interface that
includes a \texttt{run\_defense} method, which provides the current HTML, accessibility
tree, and screenshots to the defense module. The defense is then able to operate on the
observation of choice (or use multiple in tandem), to perform the desired filtering and
subsequently return a restricted observation that the base web agent then consumes.

The \sys evaluation framework provides two primary evaluation methods: \textbf{1)}
\texttt{run\_experiment()} provides an end-to-end evaluation where a web agent is tasked
to complete a user-assigned task. At the same time, attacks are injected onto the page
dynamically, and the defense restricts observations in real time. \textbf{2)}
\texttt{defense\_harness()} provides a more fine-grained evaluation of the defense,
enabling more precise diagnosis of which defense features work and which ones struggle.
This method uses
successfully completed tasks from a reference end-to-end evaluation (where neither attack
nor defense is enabled) as a ground truth. It then post-hoc evaluates the defense on each
step that the agent took in the ground truth and determines if the defense made the
action impossible through restrictions.

While we acknowledge that there may be multiple ways to complete a trajectory, and that
this ground truth may not be the shortest path to the solution, this defense harness
provides a lower-bound metric for defense utility. Moreover, it can be used
to specifically diagnose which stage of a defense is over-restricting the observation,
identifying what parts of the defense may require additional tuning. Additionally, as
there is some variance in the base web agent's performance from run to run, this
simulation-type evaluation controls for the randomness in the base web agent, ensuring
the only remaining non-deterministic factor is from the defense model itself. Lastly, in
cases where users may wish to fine-tune a model to be more specifically suited for these
tasks, the defense harness lends itself particularly well to reinforcement learning problems.

\subsection{Evaluation Metrics}
\label{subsec:evaluation-metrics}
We evaluate \sys across two metric families: attack effectiveness and labeling accuracy.
These metrics capture both security behavior and practical usability, measuring
\sys's protective capabilities against \xsp attacks.

\subsubsection{Attack Metrics} We define three metrics: task success, attack success, and
defense success.

\noindent\textbf{Task Success Rate (TSR).} Defined by WebArena for each benchmark and
computed using a combination of programmatic checks, LLM-as-a-judge, and string
matching~\cite{zhou2024webarena}.

\noindent\textbf{Attack Success Rate (ASR).} An attack is successful if the agent
interacts with injected malicious content, e.g., clicking a malicious link, or filling an
injected form; note that one-click browser exploits still appear in the
wild~\cite{GoogleTAG2024,OCearbhaillMarczak2022}.

\noindent\textbf{Defense Success Rate (DSR).} A defense step succeeds when \sys removes
or downgrades the attack to read-only where applicable. A failed attack does not imply a
successful defense, and vice versa.

\subsubsection{\sys Labeling Metrics}
\label{subsec:labeling-metrics}
We use three allowed-element labeling metrics: precision, recall, and F1. These
metrics compare a model's labeling against human-expert annotations over the DOM
elements that remain available to the agent after \sys's security policy and mechanical
confinement has been enforced.

\noindent\textbf{Allowed-element set.}
Let $A_{\text{human}}$ and $A_{\text{model}}$ be the sets of elements admitted by
\sys's security policy under the expert's and the model's labels, respectively.
Comparing the two sets isolates the elements admitted under one labeling but not the
other.

\noindent\textbf{Allowed-element Precision.}
The fraction of model-admitted elements the expert also admits:
$\mathrm{Prec} =
|A_{\text{model}} \cap A_{\text{human}}| / |A_{\text{model}}|$.
Low precision indicates over-permissiveness; it weakens security by admitting
elements the expert would reject.

\noindent\textbf{Allowed-element Recall.}
The fraction of elements the expert admits that the model also admits:
$\mathrm{Rec} =
|A_{\text{model}} \cap A_{\text{human}}| / |A_{\text{human}}|$.
Low recall indicates over-restrictiveness; it degrades agent utility but does not
weaken security.

\noindent\textbf{Allowed-element F1.}
The harmonic mean of precision and recall:
$F_1 =
2 \cdot \mathrm{Prec} \cdot \mathrm{Rec} / (\mathrm{Prec} + \mathrm{Rec})$.
We report all three because a defense's tolerable precision/recall trade-off is
application-specific.

\subsection{Implementation}

To evaluate \sys, we constructed an evaluation framework to implement various \xsp
attacks in an end-to-end web environment~\cite{zhou2024webarena}. Our framework is built
on the \texttt{rllm} project~\cite{rllm}, which provides an asynchronous and parallel
environment for evaluating agents. We then use BrowserGym~\cite{workarena2024BrowserGym}
to orchestrate the browser environment for the web agent, and to provide code for the
base web agent that we build upon. Under the hood, BrowserGym utilizes Playwright with
Chromium, an automated browser testing framework, to expose various core functions to the
web agent (e.g., \texttt{click()}, \texttt{fill()}, \texttt{go\_back}). To standardize
our LLM API requests and enable caching for Prismata, we deploy LiteLLM~\cite{litellm}
with Redis. However, we specifically disable caching by default and only enable it on
specific requests within \sys, whilst leaving caching off for the web agent attack logic.
Our \sys evaluation framework, accumulating 18,409 lines of code, is written in Python
and leverages SQLAlchemy and a PostgreSQL backend to store all the various web agent
experiment results (e.g., observations, actions, telemetry).

\sys itself is implemented as a Python-based defense module that can be plugged into the
evaluation framework, allowing for quick iterations on its design and logic. It stores
the entire DOM representation, along with a representation of the accessibility tree, to
facilitate mapping between the two representations. This enables both optimization
purposes (e.g., pruning non-visible branches) and provides a high degree of granularity
pertaining to the DOM. Its core function, \texttt{populate\_metadata}, is the logic that
implements Biba Parsing (\S\ref{subsec:security-guarantees}). Once invoked, it leverages an
asynchronous runtime, semaphores, and condition variables to launch multiple concurrent
requests and coordinate their responses simultaneously. The code for \sys is relatively
concise, at around 1,135 lines of code.
 \section{Evaluation}
In this section, we provide three primary evaluations of \sys. The first evaluates how
effectively \sys mitigates attacks across the dataset and compares it to the base web
agent without \sys (\S\ref{subsec:attack-evaluation}). The second evaluates how \sys
affects the baseline performance of the web agent in non-adversarial settings
(\S\ref{subsec:utility-evaluation}). Finally, we provide a more granular analysis of
the performance of \sys in its various stages, along with addressing potential issues
such as computational cost or time to execute (\S\ref{subsec:defense-evaluation}).

\subsection{Experimental Setup}
\label{subsec:experimental-setup}

We use WebArena which provides reliable and reproducible environments for web agent
evaluations~\cite{zhou2024webarena}, and has been used extensively by the web agent
community~\cite{marreedEnterpriseReadyComputerUsing2025,yangAgentOccamSimpleStrong2025,zhangWebPilotVersatileAutonomous2024,wangAgentWorkflowMemory2024,zhouProposerAgentEvaluatorPAEAutonomousSkill2024}.
WebArena is built from widely deployed open-source stacks: its GitLab instance runs
the real GitLab codebase, and its shopping, Reddit-style forum, and CMS sites are
backed by production-grade platforms populated with realistic data, giving its DOM
structures a fidelity comparable to the live web.
We execute each experiment three times and report mean values along with standard
deviations across runs (model snapshots are recorded in
Appendix~\ref{app:model_snapshots}).
For our base web agent, we first evaluated \texttt{gpt-5.4}, \texttt{gpt-5.4-mini}, and
\texttt{gpt-5.4-nano} on the full WebArena benchmark using the BrowserGym framework and
web agent, which achieved reward-weighted task success rates of 33.6\%, 30.8\%, and
16.0\%, respectively.
Because the differences between \texttt{gpt-5.4} and \texttt{gpt-5.4-mini} were modest,
we chose \texttt{gpt-5.4-mini} as our primary model for all experiments, balancing performance
with the cost and efficiency requirements of running the experiments required for this
paper. For our defense models, we selected candidates with inference cost lower than or
comparable to the cost of running the web agent itself.

\subsubsection{Validation Study}
\label{subsec:validation-study}
\sys's security rests on the underlying model correctly labeling ownership and
capabilities, since a mislabel can admit content the policy would otherwise restrict. We
therefore conduct a validation study to assess model performance on owner and capability
labeling. Using the same experimental setup, we randomly sample four observations from
each WebArena site, drawn from our baseline experiments regardless of task success. For
each observation, we replace the model labeler with an expert human annotator
knowledgeable in web security and development, while maintaining all other cascading
policy and enforcement settings. We then compare these human annotations against model predictions.
Each observation required approximately one hour to annotate, totaling approximately 20
hours of annotation time. We focus on benign observations as the \xsp threat model
assumes malicious content is injected into user or external fields; accurate labeling
of these fields helps \sys restrict them. We use DSR to measure how
effectively \sys mitigates attacks end to end, and use allowed-element precision, recall,
and F1 to measure how closely model-derived filtering matches expert annotations.

\begin{table}
  \caption{Attack Evaluation with \sys enabled (\CIRCLE) and without it enabled (\Circle).}
  \label{table:attacks}
  \begin{tabularx}{\linewidth}{l|>{\centering\arraybackslash}X|>{\centering\arraybackslash}X>{\centering\arraybackslash}X|>{\centering\arraybackslash}X>{\centering\arraybackslash}X|>{\centering\arraybackslash}X>{\centering\arraybackslash}X}
    \toprule
    Attack &  & TSR & $\sigma$ & ASR & $\sigma$ & DSR & $\sigma$ \\
    \midrule
    \multirow[c]{2}{*}{\footnotesize{Baseline}} & \Circle & \footnotesize{29.9} &
    \footnotesize{1.4} & -- & -- & -- & -- \\
    \footnotesize{} & \CIRCLE & \footnotesize{26.6} & \footnotesize{0.3} & -- & -- & -- & -- \\
    \midrule
    \multirow[c]{2}{*}{\footnotesize{Shortcut}} & \Circle & \footnotesize{11.2} &
    \footnotesize{0.9} & \footnotesize{63.9} & \footnotesize{2.0} & -- & -- \\
    \footnotesize{} & \CIRCLE & \footnotesize{24.4} & \footnotesize{1.1} &
    \footnotesize{2.1} & \footnotesize{0.4} & \footnotesize{88.4} & \footnotesize{2.0} \\
    \addlinespace
    \multirow[c]{2}{*}{\footnotesize{Completion}} & \Circle & \footnotesize{2.3} &
    \footnotesize{0.3} & \footnotesize{92.5} & \footnotesize{0.7} & -- & -- \\
    \footnotesize{} & \CIRCLE & \footnotesize{24.8} & \footnotesize{0.8} &
    \footnotesize{0.0} & \footnotesize{0.0} & \footnotesize{95.8} & \footnotesize{2.7} \\
    \addlinespace
    \multirow[c]{2}{*}{\footnotesize{Ignore}} & \Circle & \footnotesize{0.0} &
    \footnotesize{0.0} & \footnotesize{100.0} & \footnotesize{0.0} & -- & -- \\
    \footnotesize{} & \CIRCLE & \footnotesize{19.9} & \footnotesize{0.7} &
    \footnotesize{0.0} & \footnotesize{0.0} & \footnotesize{95.1} & \footnotesize{5.0} \\
    \bottomrule
  \end{tabularx}
\end{table}
 \begin{table}[t]
  \centering
  \footnotesize
  \caption{WASP and adaptive attack evaluations.}
  \label{table:wasp-attacks}
  \setlength{\tabcolsep}{2.8pt}
  \begin{minipage}[t]{\linewidth}
    \begin{minipage}[t]{0.51\linewidth}
      \centering
      \textbf{WASP}

      \begin{tabularx}{\linewidth}{@{}>{\raggedright\arraybackslash}Xccc@{}}
        \toprule
        Setting & \sys & Goal 0 & Goal 1 \\
        \midrule
        E2E & \Circle & 1/21 & 0/21 \\
        Int. & \Circle & 1/21 & 0/21 \\
        E2E & \CIRCLE & 0/21 & 0/21 \\
        Int. & \CIRCLE & 0/21 & 0/21 \\
        \bottomrule
      \end{tabularx}
    \end{minipage}
    \hfill
    \begin{minipage}[t]{0.45\linewidth}
      \centering
      \textbf{Adaptive WA20}

      \begin{tabularx}{\linewidth}{@{}>{\raggedright\arraybackslash}Xcc@{}}
        \toprule
        Attack & \sys & ASR \\
        \midrule
        Shortcut & \Circle & 100.0 \\
        Shortcut & \CIRCLE & 0.0 \\
        Completion & \Circle & 65.0 \\
        Completion & \CIRCLE & 0.0 \\
        Ignore & \Circle & 85.0 \\
        Ignore & \CIRCLE & 0.0 \\
        \bottomrule
      \end{tabularx}
    \end{minipage}
  \end{minipage}
\end{table}

\subsection{Attack Evaluation}
\label{subsec:attack-evaluation}

Our adversarial evaluation of \sys is presented in Table~\ref{table:attacks}, which
includes experiments with and without \sys enabled, along with WASP stress tests
(Table~\ref{table:wasp-attacks})~\cite{evtimovWASPBenchmarkingWeb2025}.
Table~\ref{table:attacks} also
includes a benign baseline to quantify the performance cost in non-adversarial
conditions, which we refer to as {\em utility}. Across all adversarial settings, \sys
increases the average task success rate (TSR) from 4.5\% to 23.0\% (+18.5
percentage points; $\approx$5.1$\times$ relative), while reducing the average attack
success rate (ASR) from 85.5\% to 0.7\% compared to the baseline without \sys.
Averaged across attacks, this corresponds to $\Delta$ASR $=-84.8$ points,
$\Delta$TSR $=+18.5$ points, and an average defense success rate (DSR) of $93.1\%$.
Under benign conditions, \sys preserves most baseline utility: the baseline TSR of
29.9\% without \sys is 26.6\% with \sys enabled.

These evaluations demonstrate that \sys is particularly effective at mitigating attacks
that convince the agent to interact with adversarial content (i.e., Shortcut),
complete the task prematurely (i.e., Completion), or distract it with ignorable content
(i.e., Ignore). For Completion and Ignore, \sys drives ASR to 0.0\% while increasing
TSR from 2.3\% and 0.0\% to 24.8\% and 19.9\%, respectively. For Shortcut, \sys
reduces ASR from 63.9\% to 2.1\% while raising TSR from 11.2\% to 24.4\%. Although
TSR under attack remains below the benign baseline, \sys consistently makes
otherwise-vulnerable trajectories both safer and more likely to complete the user's task.
Table~\ref{table:wasp-attacks} separately reports the WASP and adaptive attack
stress tests: \sys eliminates end-to-end attack success on the adaptive WA20 attacks
and blocks WASP plain-goal attack success in both goal settings. The WASP email change
task is a direct test of \xsp: it attempts to use the site's benign account-management
feature to change the user's email to \texttt{test@footest.com}. Without \sys, the WASP
plain-goal attack succeeds once in the Goal~0 setting; with \sys enabled, that success is
blocked. \sys is designed to confine that site feature when it falls outside the user's
task.

\noindent\textbf{ASR and DSR are not complements.} A successful attack always counts as a
defense failure, but the converse does not hold: an attack can fail even when \sys never
removes or downgrades it, because the agent sometimes ignores the injected content on its
own. We report both because ASR measures whether the agent was compromised, while DSR
measures whether \sys actively neutralized the attack
(\S\ref{subsec:evaluation-metrics}).

\subsection{\sys Labeling Evaluation}
\label{subsec:labeling-evaluation}

\begin{figure}[t]
  \centering
  \includegraphics[width=\linewidth]{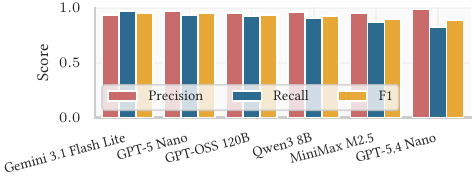}
  \caption[Cross-model allowed-element precision, recall, and F1.]{Allowed-element
    precision, recall, and F1 for the six evaluated labelers, averaged over the
    20 paired observations. Each metric compares the set of elements that survive
    \sys's defense under the model's labeling against the set that survives under
  the expert's.}
  \label{fig:allowed-element-prf-models}
\end{figure}

\begin{figure}[t]
  \centering
  \includegraphics[width=\linewidth]{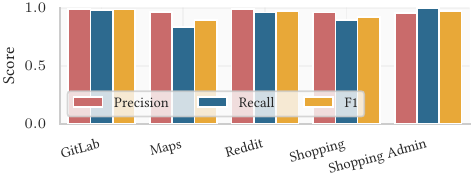}
  \caption[Per-site allowed-element metrics for GPT-5.4-nano.]{Per-site
    allowed-element precision, recall, and F1 for GPT-5.4-nano. Precision holds at
  or above $95\%$ on every site; the lowest recall occurs on Maps.}
  \label{fig:allowed-element-prf-sites}
\end{figure}

Our validation study replaces Biba parsing's ownership and capability labeling with
expert annotations. We compare the resulting allowed-element sets using the metrics
defined in Sec.~\ref{subsec:evaluation-metrics}. We sampled 20 WebArena observations
spanning five domains: GitLab, Maps, Reddit, Shopping, and Shopping Admin. Each
observation is labeled once by a domain expert and once by each candidate model.
From every labeling, we derive the set of elements that survive \sys's filtering
process and report the precision, recall, and F1 of the model's allowed-element set
against the expert's. We hold the rest of the defense stack fixed so the metrics isolate
how well each model captures the labeling decisions that \sys's filtering depends on.

Fig.~\ref{fig:allowed-element-prf-models} summarizes allowed-element precision,
recall, and F1 across the six evaluated models. Across both proprietary and
open-source models, we observe consistently high precision: every evaluated model
exceeds $93\%$ precision and $87\%$ F1, with the strongest configurations clustering
between $95\%$ and $99\%$ precision. Since over-permissive labels are the failure
mode that weakens \sys's integrity guarantee, we prioritize precision when comparing
models. Under that criterion, GPT-5.4-nano achieves the highest precision
($98.69\%$), but at the cost of the lowest recall ($82.69\%$). Gemini~3~Flash
achieves the highest F1 at $95.14\%$, while GPT-5.4-mini provides a strong balance
between precision and recall. The open-source models, GPT-OSS-120B, Qwen3-8B, and
MiniMax-M2.5, also maintain high
precision, but their lower recall suggests more conservative labeling: they tend to
exclude benign elements more often than they admit elements the expert would reject.

Fig.~\ref{fig:allowed-element-prf-sites} shows the per-site breakdown for
GPT-5.4-nano. Precision remains at or above $95\%$ on every site, and F1 remains above
$89\%$ throughout. The only meaningful recall dip occurs on Maps, where the smaller
annotated DOM makes each single-element disagreement have a larger effect on the
aggregate score. These results indicate that the models powering Biba parsing can
provide labels precise enough for \sys to enforce its policy while preserving
most benign elements. They also suggest that remaining precision gains depend more
on base-model quality than on architectural changes to \sys.

Two caveats are worth noting. First, allowed-element F1 averages over many elements,
so it does not capture worst-case safety. A $95\%$ F1 still leaves room for a single
mislabel on a critical element to allow an attack-relevant element through, although
other parts of \sys, such as capability restriction, may still contain the attack.
Second, the human annotation pass is subject to reviewer fatigue and possible human
error, so we capped the study at four observations per site to keep the labeling
effort manageable while preserving diversity across page archetypes.

\subsection{\sys Utility Evaluation}
\label{subsec:utility-evaluation}
\label{subsec:defense-evaluation}

To understand the performance impact imposed by \sys on the web agent beyond TSR, we
directly compare the benign WebArena baseline runs with and without \sys. These runs
show a small task-success-rate difference (mean difference $=3.3$ percentage points),
indicating that \sys preserves most of the agent's baseline utility while active.

\begin{figure}[t]
  \centering
  \includegraphics[width=\linewidth]{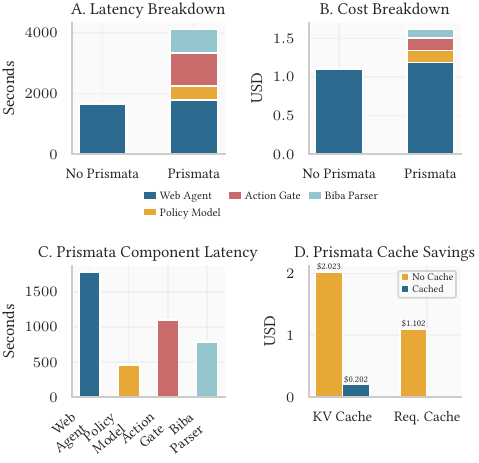}
  \caption{\textbf{Deployment cost and caching.} Panels A and B compare end-to-end
    latency and cached dollar cost for matched WA20 runs with and without \sys. Panels C
    and D isolate \sys runs, showing component latency for the web agent, policy model,
    action gate, and Biba parser, along with the cost avoided by KV and request-level
  caching.}
  \label{fig:cache-hit-rate}
\end{figure}

\sys does, however, rely on a significant number of requests in order to properly
annotate the HTML DOM. To minimize the latency and cost of these requests, we measured
the effectiveness of our caching layer~\cite{litellm}. Fig.~\ref{fig:cache-hit-rate}
summarizes this deployment overhead with four panels: the end-to-end latency breakdown,
the end-to-end cost breakdown, the per-component latency of \sys, and the caching savings
from KV and request-level reuse. In the matched WA20 runs, adding \sys increases the
mean web-agent request-duration contribution from 1,649.8s to 1,781.4s, while adding
459.0s for the policy model, 1,098.2s for action gating, and 789.6s for Biba parsing.
The simulated cached cost is \$1.105 without \sys and \$1.623 with \sys across the
20-task runs. Caching avoids \$1.821 through KV reuse and \$1.102 through request-level
reuse. Thus, \sys's extra inference cost is concentrated in reusable DOM-labeling work, so both
cache layers are essential to keeping \sys inexpensive.

  \section{Discussion \& Related Work}

In this section, we will discuss approaches from prior work in approaches to agent
security and web agent-specific defenses. We will then further contextualize how \xsp
fits into the broader security landscape as an attack class. We also expand on the
parallels between \xsp and XSS, and how the fine-grain security framework we establish to
dissect web pages in some ways mirrors prevention for XSS.

\subsection{Related Work}
\label{subsec:related-work}

In our review, we've found approaches to agent security and specifically preventing
prompt injection have often taken either one of two routes: \textbf{1)} model-based
approaches~\cite{wallaceInstructionHierarchyTraining2024,liuDataSentinelGameTheoreticDetection2025},
which are typically evaluated empirically, and \textbf{2)} systems security-based
approaches~\cite{debenedettiDefeatingPromptInjections2025,wuSystemLevelDefenseIndirect2024a,shiProgentProgrammablePrivilege2025,beurer-kellnerDesignPatternsSecuring2025,tsaiContextualAgentSecurity2025a},
that can provide more mechanical constraints and protections against attacks.

Model-based approaches to preventing prompt-injection have introduced significant steps
forward for agentic
security~\cite{wallaceInstructionHierarchyTraining2024,liuDataSentinelGameTheoreticDetection2025,liuFormalizingBenchmarkingPrompt2024a,xiangGuardAgentSafeguardLLM2025,liuPromptInjectionAttack2024,debenedetti2024agentdojo}.
However, empirically-driven approaches can also lack the ability to provide more formal
guarantees inherited by the more traditional security approaches. Although model-based
approaches have been proposed in the web agent security
landscape~\cite{zhengWebGuardBuildingGeneralizable2025,yangInContextDefenseComputer2025a},
these approaches can be more difficult to generalize, and can be more susceptible to
attacks against \xsp-like attacks~\cite{evtimovWASPBenchmarkingWeb2025}.

Under the system security umbrella, prior work has explored defenses involving
information flow control~\cite{costaSecuringAIAgents2025b, zhongRTBASDefendingLLM2025,
wuSystemLevelDefenseIndirect2024a, siddiquiPermissiveInformationFlowAnalysis2025}, for
instance. These approaches have become an increasingly popular technique used to provide
security guarantees in agentic systems that connect various systems via tool-calling.
These systems benefit significantly from rich ground-truth metadata features inherent to
the tools defined (e.g., \texttt{send\_email(recipient, subject, body)}), allowing
external security frameworks to detect information flowing in potentially harmful
directions. Unfortunately, the web agent's actions are contextually uninformative (e.g.,
\texttt{click(element=101)}). These non-descriptive actions make it difficult to assess,
with any real granularity, the security implications of actions without understanding the
broader browsing context. The only parallel ground-truth metadata, in these cases, is the
domain that the agent accesses. Unfortunately, domain-level granularity is likely
insufficient for an agentic web browser. Recalling our example, even generally trusted
domains (i.e., shopping.com) often include user content (e.g., product reviews).
Moreover, as with XSS, compromised agents may leverage site features to exfiltrate
information to attackers, even without leaving the domain (e.g., direct messages, reviews, etc.).

Specifically, CaMeL~\cite{debenedettiDefeatingPromptInjections2025} achieves
complete-mediation guarantees by pre-planning over structured tool APIs whose arguments
carry semantic ground truth and are, crucially, not data-dependent. However, every web
browsing task is inherently data-dependent: an action like \texttt{click(id)} carries no
security meaning in isolation. The same action might post a public comment, send a direct
message, or reset a password; it is the site, not the action, that decides what the click
actually does. As such, CaMeL-style pre-planning cannot be lifted directly into the web
agent setting without first solving the web entanglement problem.

Another novel complexity of the web browsing context is the general uncertainty of the
actions required to complete a task (especially prior to beginning the task), and a need
to, at times, utilize untrusted information to make decisions. While many previous
agentic defense frameworks leverage security models to plan the potential tools and
permissions required before beginning a task \cite{balunovicAIAgentsFormal2024,
debenedettiDefeatingPromptInjections2025, costaSecuringAIAgents2025b,
wuSystemLevelDefenseIndirect2024a}, web browsing is a fundamentally open-ended
interface, typically featuring multiple ways to do the same thing. Some
work~\cite{tsaiContextualAgentSecurity2025a} has begun to explore solutions that leverage
context to enable just-in-time security policies across open-ended tasks, balancing both
utility with fine-grained controls.

Dual LLM architectures for secure AI systems have been proposed
~\cite{Willison2023DualLLMPattern} and implemented
recently~\cite{debenedettiDefeatingPromptInjections2025,
beurer-kellnerDesignPatternsSecuring2025, costaSecuringAIAgents2025b}. Originally
inspired by the dual LLM approach, \sys is designed with a finer-grained separation and
strict notions of authentication prior to authorization. Dual-LLM designs
typically assume structured tool APIs in which actions and data are clearly typed,
allowing a quarantined model to pass bounded variables up to a privileged planner. Web
agents violate this assumption: their action vocabulary is abstract
(\texttt{click(elementId=\dots)}, \texttt{fill(\dots)}) and each argument is a
reference into the untrusted DOM, so deciding what to act on requires direct exposure
to page context. A strict two-level split is therefore insufficient.
We argue that processing the
untrusted content from web environments requires an intermediary level of exposure to
untrusted content (e.g., Biba parsing). Therefore, \sys introduces this intermediary level
of exposure between quarantine and full privilege.

\subsection{XSS Preventions}
\label{subsec:xss-preventions}

Browser, framework, and site developers have collaborated over the past decade to
introduce techniques and standards to mitigate these attacks, including input
sanitization~\cite{Balzarotti2008Saner_Composing,hooimeijerFastPreciseSanitizer2011},
content security policies~\cite{mdnCSPSandbox,googleAdsCSP}, for instance.

Unfortunately, state-of-the-art mitigations for XSS can not be easily adapted to web
agent environments. In the case of user content, developers can leverage
sanitizers~\cite{Balzarotti2008Saner_Composing,hooimeijerFastPreciseSanitizer2011} to
guarantee an input does not contain executable code. With agents, LLMs have no clear
distinction between data and instructions~\cite{greshakeNotWhatYouve2023}, meaning that
any textual content may be interpreted and executed as instructions. In other words, all
user content is a potential avenue for malicious payloads, and no filter can provide
formally provable security without censoring all content.

Rigorous protections have also been developed to ensure that external content, such as
advertisements, cannot execute malicious instructions. Browser-based detection
techniques~\cite{stockPreciseClientsideProtection2014,Vogt2007XSS} and sandboxing with
iframes~\cite{mdnIframe} provide advertisements with an isolated document environment in
which to operate, and the Content-Security-Policy (CSP) sandbox
directive~\cite{mdnCSPSandbox} further restricts ad capabilities. These browser
primitives are leveraged by platforms such as Google AdSense, which implement the
SafeFrame specification~\cite{safeFrameSpec} to ensure advertisements are restricted in
their capabilities to interact with the host site~\cite{googleAdsenseSafeframe}, along
with CSP standards~\cite{googleAdsCSP}. Unfortunately, once again, web agents have no
real notion of fine-grain permissions and are unable to guarantee restricted capabilities
or privileges for any content within the LLM context.

This absence of granular controls positions \sys as a novel security framework, allowing
a web page to be subcategorized into different regions of ownership with distinctive
capabilities. In the same way that iframe directives can provide or restrict browser
capabilities in advertisements, such as \texttt{allow-downloads} or
\texttt{allow-popups}~\cite{mdnIframe}, \sys provides read-only or read-write
capabilities for the web agent.

\subsection{Limitations}
\label{subsec:limitations}

\noindent\textbf{Integrity vs. Accuracy:} A key limitation of \sys is an intentional
trade-off of integrity vs. accuracy, where \sys opts to preserve integrity over
prioritizing accuracy. More specifically, we mean that a model with access to an
element's parents and children may predict more accurate labels pertaining to each
element's purpose and context. However, our approach mechanically blocks upward
tampering and privilege escalation under the stated action-space and threat-model
assumptions.

\sys is limited by the accuracy of the underlying model used to produce outputs at the
various stages (i.e., the policy model, action gate, and Biba parsing). As we demonstrate in our
evaluation (\S\ref{subsec:utility-evaluation}), different models produce various
results in this regard. To further improve performance, it would be straightforward to
utilize a reinforcement learning pipeline to fine-tune these components. However, for our
evaluations, we found this step unnecessary, as commodity models already provide
satisfactory performance.

\noindent\textbf{Context Availability:} In the same regard, it is essential to recognize
that \sys utilizes the HTML and available attributes (e.g., class names, href, etc.) to
determine the appropriate metadata and capabilities defined. As such, there is the
potential for mislabeling, which is influenced by the site's specific HTML layout and
structure. In practice, \sys's labeling draws on multiple signals, including ancestry in
the DOM, visible headers (e.g., ``Reviews,'' ``Comments''), link targets, and other
metadata, rather than any single attribute. Labeling is inherently heuristic: our
validation study (\S\ref{subsec:validation-study}) is designed to enable comparison
across models against a security-oriented human reference, not to assert ground truth.
Attack success rate and baseline utility then serve as complementary end-to-end
proxies, since overly permissive labeling would raise ASR while overly restrictive
labeling would degrade TSR. This posture is consistent with other application-layer
defenses (e.g., WAFs, IDS, spam filters) that operate under imperfect signals yet
remain valuable within a defense-in-depth strategy.
In practice, several widespread conventions produce the structural cues that \sys
relies on: UI design conventions place headings and labels before the content they
describe; accessibility standards such as WCAG require section headers before their
content; and component-based development practices, including semantic HTML and
methodologies like BEM, produce descriptive wrapper elements around dynamically injected
content. These conventions are reinforced by economic incentives: search engines both
reward and rely on semantic structure and accessibility compliance to parse and rank
content.
Collaboration with site developers to integrate standardized labels, such as
the ones used in this work, would further alleviate the issue. As web agents continue
to gain prominence, it is not impossible that these types of websites or standards could
be adopted, especially considering the standards adopted to prevent
XSS~\cite{safeFrameSpec,mdnIframe}.

\noindent\textbf{Captured vs. Live DOM:} \sys operates on the same serialized DOM
snapshot that is handed to the underlying agent, so any drift between the captured
observation and the live page at execution time affects the agent and \sys equally.
This gap is a fundamental limitation of LLM-driven web agents, which must serialize
the page into text before reasoning about it, rather than an artifact introduced by
\sys.

\noindent\textbf{Non-Textual Attacks:}
We acknowledge that \sys does not cover all domains of attacks web agents will encounter
in multimodal environments. Since we evaluate web agents using accessibility trees as the
primary input, we recognize that multimodal agents will also utilize non-textual
modalities such as visual and aural inputs. Therefore, gradient-based image perturbations
\cite{luoImageWorth10002023}, non-textual jailbreaking attacks
\cite{gengConInstructionUniversal2025, liImagesAreAchilles2025}, and multimodal attacks
\cite{wangWhiteboxMultimodalJailbreaks2024a} are still exploitable for web agents using
\sys. However, \sys's core security principles may be applied to develop defenses
targeting non-textual modalities, enabling future system-level defenses that provide
comprehensive protection for multimodal agents.

\noindent\textbf{Hallucinations:} Additionally, hallucinations are currently an issue
that plagues LLMs in general~\cite{trustllm}. Structured outputs that define predefined
output values help to constrain the output hallucination space, ensuring all outputs are
sampled from a limited set of possibilities. Furthermore, we can ensure that malicious
content will be constrained by permissions that it cannot influence, as shown in our
Prismata Security argument (\S\ref{subsec:security-guarantees}).

\noindent\textbf{High-Risk Tasks:} Lastly, we acknowledge that some tasks are likely to
require all types of content (e.g., user and external), with all capabilities provided
to untrusted content. In these situations, we propose that future work explore solutions
that delegate significant untrusted work to quarantined web agents in a manner similar to
what has been done for tool calling with quarantined
models~\cite{debenedettiDefeatingPromptInjections2025,beurer-kellnerDesignPatternsSecuring2025}.

\noindent\textbf{Agent architectures.} Some web agents have begun to explore
multi-agent designs. The filtered observation produced by \sys could theoretically be
ingested by subagents in these architectures, as well as by single-agent architectures.
 \section{Conclusion}

In this work, we present \sys, a novel contextual least-privilege defense for securing
web agents from
\xsp attacks, inspired by the 1977 classical Biba integrity
model~\cite{biba1977integrity}. \sys proposes a fine-grain security framework to dissect
pages and provide granular access control and capabilities to web agents. \sys then
interprets the web page through this framework, providing security properties that
significantly constrain \xsp attacks from conducting malicious activity through user or
external content. To provide upward integrity, we introduce Biba parsing, a novel
technique for parsing untrusted data with LLMs, which the classic Biba integrity model
once again inspires. Finally, we evaluate \sys across three web agent \xsp attack
templates, measuring a reduction in attack success rate from 85.5\% to 0.7\%, while
improving task success rate under adversarial conditions from 4.5\% to 23.0\%, all while
preserving most benign web agent utility. Lastly, our \sys evaluation framework
provides future researchers with an end-to-end framework to evaluate future \xsp attacks
and defenses.

\bibliographystyle{ACM-Reference-Format}

\appendix

\clearpage
\onecolumn

\makeatletter
\newenvironment{apxtable}[1][]{\par\medskip
\begin{center}\def\@captype{table}}{
\end{center}\par\medskip}
\let\table\apxtable
\let\endtable\endapxtable
\expandafter\let\csname table*\endcsname\apxtable
\expandafter\let\csname endtable*\endcsname\endapxtable
\makeatother

\refstepcounter{section}
\label{app:action-space}

\begin{table*}[t]
  \centering
  \small
  \begin{tabularx}{\textwidth}{@{}llX@{}}
    \toprule
    \textbf{Action} & \textbf{Parameters} & \textbf{Description} \\
    \midrule
    \multicolumn{3}{@{}l}{\textit{Element-targeting actions}} \\
    \addlinespace[2pt]
    \texttt{click}          & \texttt{bid, button, modifiers}        & Click an element
    (left/middle/right; optional modifier keys). \\
    \texttt{dblclick}       & \texttt{bid, button, modifiers}        & Double-click an element. \\
    \texttt{fill}           & \texttt{bid, value}                    & Fill an input,
    textarea, or contenteditable element. \\
    \texttt{clear}          & \texttt{bid}                           & Clear an input field. \\
    \texttt{select\_option} & \texttt{bid, options}                  & Select one or more
    values in a \texttt{<select>} element. \\
    \texttt{hover}          & \texttt{bid}                           & Hover over an element. \\
    \texttt{press}          & \texttt{bid, key\_comb}                & Focus an element
    and press a key combination. \\
    \texttt{focus}          & \texttt{bid}                           & Focus an element. \\
    \texttt{drag\_and\_drop}& \texttt{from\_bid, to\_bid}            & Drag one element
    onto another. \\
    \addlinespace[4pt]
    \multicolumn{3}{@{}l}{\textit{Navigation actions}} \\
    \addlinespace[2pt]
    \texttt{tab\_focus}     & \texttt{index}                         & Switch to a
    browser tab by index. \\
    \texttt{new\_tab}       &                                        & Open a new browser tab. \\
    \texttt{go\_back}       &                                        & Navigate back in
    browser history. \\
    \texttt{go\_forward}    &                                        & Navigate forward
    in browser history. \\
    \texttt{goto}$^\dagger$ & \texttt{url}                           & Navigate directly
    to a URL. \\
    \addlinespace[4pt]
    \multicolumn{3}{@{}l}{\textit{Control actions}} \\
    \addlinespace[2pt]
    \texttt{noop}           & \texttt{wait\_ms}                      & Wait for a
    specified duration (default 1000\,ms). \\
    \texttt{send\_msg\_to\_user} & \texttt{text}                     & Finish the task
    and respond to the user. \\
    \texttt{report\_infeasible}  & \texttt{reason}                   & Report that the
    task cannot be completed. \\
    \bottomrule
  \end{tabularx}
  \vspace{2pt}
  {\footnotesize $^\dagger$Exposed only in the WASP action set, not the standard WebArena
  action set.}
  \caption{Web agent action space. Element-targeting actions identify their target by
  BrowserGym element id (\texttt{bid}). The agent emits exactly one action per step.}
  \label{tab:action-space}
\end{table*}

\refstepcounter{section}
\label{app:model_snapshots}

\begin{table}[t]
  \centering
  \caption{Foundation models used.}
  \label{tab:empirical-foundations-model-snapshots}
  \begin{tabular}{lll}
    \toprule
    model & snapshot\_slug & provider \\
    \midrule
    gpt-5.4 & gpt-5.4-2026-03-05 & OpenAI \\
    gpt-5.4-mini & gpt-5.4-mini-2026-03-17 & OpenAI \\
    gpt-5.4-nano & gpt-5.4-nano-2026-03-17 & OpenAI \\
    gemini-3-flash-preview & no public snapshot & Google \\
    gpt-oss-20b & no public snapshot & OpenAI \\
    \bottomrule
  \end{tabular}
\end{table}
 
\begin{table}[t]
  \centering
  \caption{Evaluation model snapshots used.}
  \label{tab:evaluation-model-snapshots}
  \begin{tabular}{lll}
    \toprule
    model & snapshot\_slug & provider \\
    \midrule
    gpt-5.4 & gpt-5.4-2026-03-05 & OpenAI \\
    gpt-5.4-mini & gpt-5.4-mini-2026-03-17 & OpenAI \\
    gpt-5.4-nano & gpt-5.4-nano-2026-03-17 & OpenAI \\
    \bottomrule
  \end{tabular}
\end{table}

\refstepcounter{section}
\label{app:labeler_details}

\end{document}